\documentclass{elsart}
\usepackage{amsfonts}
\usepackage[dvips]{graphics}

\newcommand{\stacksymb}[3][]{#2\limits_{#3}^{#1}}
\newcommand{\limgo}{\mathop{\longrightarrow}}
\newcommand{\asym}{\mathop{\asymp}}
\newcommand{\Prob}{\mbox{\rm Prob}}
\journal{PHYSICA D}
\date{ }

\begin{document}

\begin{frontmatter}
\title{Large deviations from the thermodynamic limit in
       globally coupled maps}
\author{Andreas Hamm}
\address{Fachbereich Physik, Universit\"{a}t GH Essen, 45117 Essen, Germany\\
         E-Mail: Andreas.Hamm@uni-essen.de}
\begin{abstract}
Systems of a large number $N$ of globally coupled maps have become popular as
a relatively simple prototype of high-dimensional dynamics, showing
many interesting and typical phenomena like synchronisation, cluster formation
and multistability, and having potential applications in systems like
Josephson junction arrays or in biophysical models.

There exists a wealth of numerical investigations of globally coupled maps.
While much progress has been made in the explanation of the macroscopic
behaviour of such systems in the limit $N\to\infty$, there is still need for a
sound theory about the asymptotic behaviour of finite-$N$ systems as $N$
approaches infinity. 

This article introduces a method by which it is possible to obtain
asymptotic estimates for long-term deviations from the thermodynamic limit
behaviour. This method is based upon the concept of quasipotentials,
originally developed by Frei\-dlin, Wentzell, and others for describing the
influence of small random perturbations on the long-term behaviour of
dynamical systems.

The problems of explicitly computing quasipotentials in the present
context and potential approximation schemes are discussed. All the concepts
described in this article are illustrated with a simple example. 
\end{abstract}
\begin{keyword}
Globally coupled systems, mean-field dynamics, invariant measures,
quasipotentials, large fluctuations, relative entropy\\
{\em PACS: 05.40.+j, 02.50.-r, 05.45.+b}
\end{keyword}
\end{frontmatter}
\newpage

\section{Introduction}
\label{se:1}
The last decades have seen a very successful development of new concepts and
methods for the investigation of low-dimensional dynamical systems. 
It is obviously worth while to  
draw inspiration from this low dimensional experience when studying the
much more difficult and
much less explored field of high and infinite dimensional dynamics,
especially in the context of spatially extended systems.

This strategy has motivated the construction of certain models of
high dimensional dynamical systems which are particularly well suited for
an analysis in the tradition of what has been done for low dimensional
systems. One class of such models has been termed {\em coupled map lattices}
\cite{CML}.
Here, time as well as the spatial aspect of the system are discretised.  
The most commonly studied coupling form in such systems is a
nearest-neighbour coupling. There exists a large number of numerical and
phenomenological investigations about such systems, and they have been very
useful for studying typical effects appearing in extended systems as well as
for formulating theoretical concepts appropriate for the characterisation of
such systems.

However, for certain explicit calculations it is attractive to look at a
different form of coupling, which is not local but global, and which gives each
subsystem the same influence on every other subsystem 
\cite{K:D41}. Obviously, mean-field
type approximations of models with local coupling lead to such a global
coupling. These approximations are usually good for long range interactions
\cite{SBAL}
or for nearest-neighbour couplings on high-dimensional lattices
\cite{ChM92,ChM96}, but all
spatial characteristics of the system loose their meaning since all subsystems
are treated equally independent on where they are.

On the other hand, global coupling does not only appear as an approximation;
there are many 
applications where one is interested in a global coupling of the subsystems
right from the beginning. Typically, this is the case in systems for which a
macroscopic (i.e. global) conserved quantity exists, which mediates the global
coupling. Prominent examples of globally coupled systems (in the form of
globally coupled differential equations) are arrays of Josephson junctions
\cite{AJJ}, 
laser arrays \cite{ALA}, multi-mode lasers \cite{AML}, certain chemical
reactions \cite{ACR}, interacting
biological clocks \cite{ABC}, and models of neural activity \cite{ANA}.

In this paper we study the discrete-time version of globally coupled systems:
globally coupled maps.

Before formulating in the next section a more general definition of the type of
systems which we want to consider, we give a typical example: a system of $N$
globally coupled tent maps for the discrete time (index $t$) evolution of the
subsystem states 
\begin{equation}
\label{eq:1-10}
 x^{(i)}_{t+1} = f_{a}(x^{(i)}_{t}) 
               + \frac{\kappa}{N} \sum_{j=1}^{N} f_{a}(x^{(j)}_{t}),
               \qquad\qquad i=1,\dots,N
\end{equation}
with
\begin{equation}
\label{eq:1-20}
  f_{a}(x) = \frac{a-1}{2} - a|x|, \qquad\qquad 1<a\leq 2 .
\end{equation}
Here, the parameter $\kappa$ has the meaning of a coupling strength, and $a$
characterises the slope of the tent.

A very rich variety of dynamical phenomena in the long term has been
observed in this system (or in a system of globally coupled logistic maps), 
caused by the competition between a possibly chaotic
local dynamics $f_{a}$ and the ordering influence of the coupling term
\cite{K:D41,K:L65,K:D55,K:D86,PSC,Vul,SK:D124,Vul2,JGa}.
Depending on the values of the parameters $\kappa$ and $a$, the system
can show complete synchronisation of all subsystems, partial synchronisation
with the formation of synchronised clusters, or turbulent, seemingly
uncorrelated movement of all subsystems. In addition, multi-stability, often
with infinitely many attractors, is a common feature.

These phenomena seem to be typical for all globally coupled systems.

A technical merit of global coupling is that it is relatively easy to analyse
the thermodynamic limit $N\to\infty$ of the time evolution of globally
averaged quantities like the mean field
\begin{equation}
\label{eq:1-30}
  h_{t} = \frac{1}{N} \sum_{j=1}^{N} f_{a}(x^{(j)}_{t}) .
\end{equation}

In the thermodynamic limit, the time evolution of the mean-field can be
obtained from the so called {\em nonlinear Frobenius-Perron equation}
\cite{K:L65,K:D55,PSC,PKa,PKb}, an
equation for probability densities $\rho_{t}(x)$ on the subsystem state space:
\begin{equation}
\label{eq:1-40}
  \rho_{t+1}(y) = \int \delta(y-f_{a}(x)-\kappa h[\rho_{t}])
                       \rho_{t}(x) \d{x} ,
\end{equation}
where
\begin{equation}
\label{eq:1-50}
  h[\rho] := \int f_{a}(x)\rho(x)\d{x} .
\end{equation}

This equation has to be solved self-consistently. Here, the tent map
(\ref{eq:1-20}) brings a special advantage: Because of the piecewise
linearity of the map it is possible to approximate the densities
$\rho_{t}(x)$ by piecewise constant functions, and this leads to a very
accurate numerical procedure to follow the long-term dynamics of equations
(\ref{eq:1-40}) and (\ref{eq:1-50}),
see e.~g. \cite{Mor,NKPRE}. While for a truly turbulent, uncorrelated
behaviour of the subsystems one would expect to find a fixed point attractor
for the dynamics of the densities, much more complicated attractors have been
observed, hinting at some nontrivial, 'hidden' correlations in the system even
in cases without any partial synchronisation. One example for such a
non-trivial attractor is shown in Fig.~1, where $h_{t+1}$ is
plotted versus $h_{t}$ for $a=1.65$ and $\kappa=0.27$. Looking at this figure 
there appears to be a low-dimensional quasiperiodic
attractor of the nonlinear Frobenius Perron dynamics. (More detailed
calculations have shown that there is a fine structure in the time delay
mean-field representation which cannot be resolved in Fig.~1.)

During the last few years the nonlinear Frobenius Perron equation has been the
subject of numerous numerical and theoretical investigations
\cite{GriH,Jus1,Jus2,Jus3,Jarv,ErP1,ErP2,CMor,NKpp}.

However, the temporal behaviour of the mean-field can be predicted correctly
by the nonlinear Frobenius Perron dynamics only in the limit $N\to\infty$.
For large but finite $N$ there are clear deviations of the mean-field from its
limit behaviour. It is worth while to study in more detail the $N\to\infty$
asymptotics of the long-term behaviour of the mean-field because of the
following reasons:

a) Numerical simulations of a system of globally coupled maps are bound to
use a finite number $N$ of maps. It is important to be able to estimate how
large one has to choose $N$ in such simulations in order to recognise the 
details of the Frobenius Perron dynamics. 

b) Globally coupled systems which are motivated by realistic examples consist
of a finite number of subsystems. If one wants to use for their description
the results obtained in the thermodynamic limit one has to understand in
detail  how the limit situation is approached as the number of subsystems
grows. 

c) The nonlinear Frobenius Perron equation may produce features that are
sensitive to the system size in the sense that they cannot be observed for
finite $N$ at all, while other features will still be present. In order to be
able to distinguish between sensitive and insensitive features it is crucial
to understand the $N\to\infty$ asymptotics that leads to the Frobenius Perron
dynamics. 

It is easy to get a first idea about the way in which finite $N$ changes the
picture obtained in the thermodynamic limit by comparing Fig.~1
with the results of some numerical simulations with finite values of $N$.
Fig.~2 shows numerical simulations with a decreasing number of
subsystems. One can see clearly, that the details of the attractor of the
Frobenius Perron equation get washed out more and more, the fewer subsystems
are involved.

This picture resembles very much the changes in the appearance of attractors
caused by small external random perturbations; a decreasing number of
subsystems here corresponds to an increasing noise strength there. 

The idea of describing deviations from the thermodynamic limit behaviour as
stochastic perturbations is quite obvious, but it is much less easy to specify
concretely which properties these stochastic perturbations must have. Pikovsky
and Kurths \cite{PKb} gave heuristic arguments based on the central limit
theorem of how to add random perturbations to the Frobenius Perron dynamics in
order to get a proper description of numerical simulations with finite
$N$. While this method works very well for special examples it lacks a solid
theoretical justification, so that it is neither clear how widely it is
applicable nor how to improve it.

It is the aim of this article to develop an approach to understanding
long-term deviations from the thermodynamic limit in globally coupled maps 
based on {\em large deviations} methods. This means that we
concentrate on the rare but very influential events that in the case of very
large but finite $N$ render the behaviour of the coupled system perceptibly
different from what the thermodynamic limit predicts. 
The central object of this approach is a so called quasipotential. 

The notion of quasipotentials originated in the physical context from attempts
to generalise the idea of a thermodynamic potential to nonequilibrium
situations (see \cite{GraR} for a review).  Its mathematical foundations were
formulated by Freidlin and Wentzell \cite{FW}.  Quasipotentials are asymptotic
estimates --- on a logarithmic scale --- of invariant probability measures of
stochastically perturbed dynamical systems with decreasing noise
strength. They have proven to be successful for characterising the stability
and sensitivity of attractors with respect to small noise, for
estimating noise-induced escape times and transition times between
different attractors, and for deriving noise scaling laws which describe the
influence of small noise on bifurcation scenarios; a very short and
incomplete list of typical applications may help the reader to trace
recent developments in this area: \cite{HmG2,HmTG,Rei,Kau,MSt,Dyk}.

In constructing quasipotentials for the $N\to\infty$-asymptotics in globally
coupled maps, a rigorous, abstract identification of finite-$N$ effects and
random noise is established, and the above mentioned successes in applying
weak-noise quasipotentials can hopefully be repeated for large-$N$
quasipotentials. 

It should be mentioned that as a starting point for studying large deviations
in globally coupled systems we do not use pure globally coupled deterministic
maps like (\ref{eq:1-10}) but globally coupled discrete-time Markov processes
(which can be thought of as stochastic perturbations of maps). This extension
is interesting for applications and advantageous from a technical point of
view. It is interesting to note that the corresponding `noisy' version of
the nonlinear Frobenius Perron equation (touched upon in \cite{PKb,K:D55}) 
is the
discrete-time version of the nonlinear Fokker-Planck equation introduced by
Desai and Zwanzig \cite{DesZ}. A quasipotential approach to finite-$N$
deviations from the nonlinear Fokker-Planck equation has been derived by
Dawson and G\"{a}rtner \cite{DawG} and is closely related to the present
approach. However, the discrete-time theory described in this paper is clearly
conceptually less complicated and easier to use for concrete calculations. 

The plan of this article is as follows: 
In the next section we give a definition of exactly which type of coupled
systems we want to investigate. In addition, that section serves to introduce 
our notation.
 
Section \ref{se:3} is the theoretical main part of the article.
Subsection \ref{se:3}.1 deals with the time-evolution in the thermodynamic
limit. In subsection \ref{se:3}.2, deviations from the thermodynamic limit in
each single time step are studied. Subsection \ref{se:3}.3 shows how to obtain
asymptotic estimates for long-term deviations from the thermodynamic limit by
putting together the information about the single time steps and thereby 
constructing a quasipotential on a space of probability measures.
Since this `full' quasipotential is a rather indigestible object,
in subsection \ref{se:3}.4 {\em contracted quasipotentials} are introduced as a
way of extracting information about the long-term behaviour of macroscopic
variables.  

Section \ref{se:4} turns to the question, how large-$N$ quasipotentials can be
computed. In particular, in subsection \ref{se:4}.1 a discrete-time
Hamiltonian field theory with constraints is set up, emerging from a
variational principle fulfilled by the contracted quasipotentials. While it
does not seem to be possible to solve the corresponding Hamilton equations in
general, one can introduce approximations that lead to an algorithm which is
suitable for numerical computations of contracted quasipotentials close to
attractors of the dynamics of macroscopic variables. In subsection
\ref{se:4}.2 the ideas and the approximation schemes worked out in this
article are illustrated with a simple example of globally coupled linear maps
with Gaussian noise.

Finally, Section \ref{se:5} sums up the statements of the paper and indicates
in which directions ongoing and planned investigations are heading.

An Appendix gives a self-contained proof of Theorem \ref{th:2} (which is one
of the main ingredients of the method developed in this article). Its first
paragraphs can serve as an introduction into the language used for
probabilistic arguments throughout this paper; more details on the relevant
basic facts from probability theory can be found in introductory texts like
\cite{Dud,Tay}.

\section{Systems of globally coupled noisy maps}
\label{se:2}
The state of a system which consists of $N$ coupled subsystems, each being
defined on some state space $M$, can be characterised by an $N$-tuple
\begin{equation}
\label{eq:2-10}
  \mathbf{x}^{[N]} := (x^{[N](1)}, \dots , x^{[N](N)}) \in M^{N},
\end{equation}
which we call the configuration of the coupled system. Often it is not
necessary to mention the index $[N]$ explicitely in the components of 
$\mathbf{x}^{[N]}$, so that we usually write $x^{(i)}$ instead of
$x^{[N](i)}$. 

In this and the next section we assume that $M$ is a compact metric space. 
(However, the
assumption of compactness could be weakened at the cost of a much higher
technical effort in most of our considerations.) The most common case studied
in concrete examples is $M=I\subset\mathbb{R}$, a closed interval of the real 
line.

Coupling means generally that the subsystems do not follow an individual time
evolution, but that the state of any subsystem at a later time depends on the
states of other (and in the case of {\em global coupling} even of all)
subsystems at an earlier time.  We are interested in a form of global coupling
which can be incorporated in a measure theoretic way and which is sometimes
called {\em democratic coupling} since all elements have equal influence in
it.

We first define one of the central objects of our investigation, the 
{\em empirical measure} connected to a configuration
$\mathbf{x}^{[N]}$:
\begin{equation}
\label{eq:2-20}
  \mathcal{E}^{[N]}_{\mathbf{x}^{[N]}}
   := \frac{1}{N} \sum_{i=1}^{N} \delta_{x^{(i)}}
   \in \mathcal{P}(M).
\end{equation}

Here, $\mathcal{P}(M)$ denotes the set of Borel probability measures
\cite{Dud} on the
subsystem state space $M$, and $\delta_x$ for $x\in M$ is the Dirac measure
concentrated on $x$ which is defined for all Borel sets $A\in M$ by 
$\delta_x(A)=1$ if $x\in A$ and $\delta_x(A) = 0$ if $x\not\in A$. 

The concept of empirical measures has its origin in statistical estimation,
but here it is simply a means of counting how many subsystems are in which
state for a given configuration:
If $A\subset M$ and $\mathbf{x}^{[N]}$ is the state of the coupled system, then 
$\mathcal{E}^{[N]}_{\mathbf{x}^{[N]}}(A)$ gives the fraction of subsystems with
subsystem state in $A$. The density of the empirical measure (which for finite
$N$ is defined only in the distribution sense) has been called `snapshot
distribution' by other authors \cite{K:D55}.

Now we define a deterministic temporal evolution of the coupled system with
discrete time $t\in \mathbb{N}_{0}$ by
\begin{equation}
\label{eq:2-30}
  x^{(i)}_{t+1} = F(x^{(i)}_{t},\mathcal{E}^{[N]}_{\mathbf{x}^{[N]}_{t}}) .
\end{equation}
with a continuous function $F:M\times\mathcal{P}(M)\rightarrow M$. Note that
the subsystem states $x^{(i)}_{t}$ carry two indices now, the upper one for
numbering the subsystem, and the lower one for denoting the time.

When talking about continuity, we have to specify the topologies which we use.
On the state space $M$ this is of course the topology derived from the 
metric on $M$. On the space of probability measures $\mathcal{P}(M)$ we use 
(unless when mentioned otherwise) the topology of weak
convergence; a sequence of probability measures $(\mu_{n})$ is said to converge
weakly to $\mu$ if for all bounded continuous functions 
$g:M\rightarrow\mathbb{R}$:
\begin{equation}
\label{eq:2-40}
  \int_{M}g(x)\mu_{n}(\d x)
  \stacksymb{\limgo}{n\rightarrow\infty}
  \int_{M}g(x)\mu(\d x);
\end{equation}
we then write
$\mu_{n}\stacksymb[w]{\limgo}{n\rightarrow\infty}\mu$. 

Later we will implicitly make use of the fact, that there is a metric on
$\mathcal{P}(M)$ (e.~g., the so called {\em L\'{e}vy-Prohorov metric}
\cite{Dud}) which
is compatible with the topology of weak convergence,
and that compactness of $M$ implies compactness of $\mathcal{P}(M)$.

It is useful to introduce a real nonnegative parameter $\kappa$ as {\em
coupling strength}: We study continuous families 
$(F_{\kappa}:M\times\mathcal{P}(M) \rightarrow M)$ which have the 
property that $F_0$ does not depend on its second argument:
\begin{equation}
\label{eq:2-50}
  F_0(x,\mu) = f(x)
\end{equation}
where $f:M\rightarrow M$ is a continuous function.
This means that for $\kappa=0$ all subsystems evolve independently of each
other according to the map $f$ whereas for $\kappa>0$ a global coupling
term supervenes the local dynamics. In this sense the time evolution
(\ref{eq:2-30}) can be regarded as a system of globally coupled maps.

The concept of empirical measures leads to a definition of {\em global} or
{\em macroscopic} variables appropriate for globally coupled systems: A
macroscopic variable is an observable which depends on the configuration
of the coupled system only through its empirical measure. Important 
examples for $M\subset\mathbb{R}$ are the mean value
\begin{equation}
\label{eq:2-60}
  m^{[N]}(\mathbf{x}^{[N]}) := \tilde{m}(\mathcal{E}^{[N]}_{\mathbf{x}^{[N]}})
\end{equation}
with
\begin{equation}
\label{eq:2-70}
  \tilde{m}(\mu) := \int_{M} x \mu(\d x)
\end{equation}
and the mean field
\begin{equation}
\label{eq:2-80}
  h^{[N]}(\mathbf{x}^{[N]}) := \tilde{h}(\mathcal{E}^{[N]}_{\mathbf{x}^{[N]}})
\end{equation}
with
\begin{equation}
\label{eq:2-90}
  \tilde{h}(\mu) := \int_{M} f(x) \mu(\d x)
\end{equation}
of a configuration.

The abstract form (\ref{eq:2-30}) of the time evolution contains the various
concrete models of globally coupled maps 
which have been studied in the literature:

Our introductory example (\ref{eq:1-10}) would correspond to
\begin{equation}
\label{eq:2-110}
  F_{\kappa}(x,\mu) = f_{a}(x) + \kappa \tilde{h}(\mu) .
\end{equation}
Here and in the other models the local map $f_{a}$
depends on some parameter $a$ which usually changes the strength of the
nonlinear character of $f_{a}$.

However, the choice
\begin{equation}
\label{eq:2-100}
  F_{\kappa}(x,\mu) = (1-\kappa)f_{a}(x) + \kappa \tilde{h}(\mu) 
\end{equation}
used by Kaneko \cite{K:D41} has the advantage that there is a built-in
guarantee that the right hand side is in $M$ as long as $\kappa$ is in the
interval $[0,1]$.   

The coupling can act through the parameter $a$, too, like in
\begin{equation}
\label{eq:2-120}
  F_{\kappa}(x,\mu) = f_{a_{0}+\kappa\tilde{m}(\mu)}(x) ,
\end{equation}
which has been studied by Pikovsky and Kurths \cite{PKb}.

As already explained in the introduction, in this article we study
coupled systems with a stochastic time evolution, for which the
deterministic behaviour of eq. (\ref{eq:2-30}) is a limit case.
What we call a system of globally coupled noisy maps is a time
discrete Markov process $(\mathbf{X}^{[N]}_{t})$ on $M^N$ which is 
defined by its one-step transition probabilities 
that have the following form:
\begin{equation}
\label{eq:2-130}
  \Prob\{ \mathbf{X}^{[N]}_{t+1}\in A^{(1)}\times\dots\times A^{(N)} 
          | \mathbf{X}^{[N]}_{t} = \mathbf{x}^{[N]} \}
  = \prod_{i=1}^{N} Q_{F(x^{(i)},\mathcal{E}^{[N]}_{\mathbf{x}^{[N]}})} 
           (A^{(i)})
\end{equation}
with $A^{(1)},\dots,A^{(N)} \subset M$, where $(Q_x)$ is a family of 
probability measures in $\mathcal{P}(M)$, continuously parametrised by 
$x\in M$.

One can think of the time evolution (\ref{eq:2-130}) as first a deterministic
step according to (\ref{eq:2-30}) and then, independently for every
subsystem, a stochastic perturbation distributed as described by the
probability measures $Q_x$. 

A concrete example in extension of (\ref{eq:2-110}) would be
\begin{equation}
\label{eq:2-150}
  X^{(i)}_{t+1} = f(X^{(i)}_{t}) + \frac{\kappa}{N}
                                  \sum_{j=1}^{N} f(X^{(j)}_{t}) 
                  + \xi^{(i)}_{t} ,
\end{equation}
where the $\xi^{(i)}_{t}$ are independent random variables.
If $\rho_{Y}(\xi)$ is the probability density of the random variable 
$\xi^{(i)}_{t}$, which generally depends on the unperturbed state
$Y:=f(X^{(i)}_{t}) + \frac{\kappa}{N} \sum_{j=1}^{N} f(X^{(j)}_{t})$,
then 
\begin{equation}
\label{eq:2-160}
  Q_Y(A) = \int_{\{\xi:Y+\xi\in A\}} \rho_{Y}(\xi) \d\xi .
\end{equation}

With the choice $Q_x = \delta_x$ the stochastic time evolution 
(\ref{eq:2-130}) reduces to the deterministic one (\ref{eq:2-30}).

In order to interpret (\ref{eq:2-130}) as a small stochastic 
perturbation of (\ref{eq:2-30}), 
a non-negative parameter $\eta$ can be introduced as
a {\em noise strength} in families 
$(Q^{[\eta]}_x)$ with 
$Q^{[\eta]}_x \stacksymb[w]{\limgo}{\eta\rightarrow 0} \delta_x$. 

However, in this article we make no explicit use of the concept of noise 
strength; we just mention that concrete calculations
are expected to be easier for small noise strengths and that work on
the small noise limit --- again using large deviation methods --- has
started \cite{SedH}.

\section{The stochastic process of empirical measures and its thermodynamic
         limit}
\label{se:3}

The stochastic process of configurations, $\mathbf{X}^{[N]}_{t}$, 
characterised by eq. (\ref{eq:2-130}) induces a stochastic 
process of empirical measures, $(\mathcal{E}^{[N]}_{\mathbf{X}^{[N]}_{t}})$,
which is again a Markov process and therefore can be characterised by its
one-step transition probability
\begin{equation}
\label{eq:3-10}
  \Prob\{ \mathcal{E}^{[N]}_{\mathbf{X}^{[N]}_{t+1}} \in B
          | \mathcal{E}^{[N]}_{\mathbf{X}^{[N]}_{t}} = \mu \},
\end{equation}
where $B$ is a Borel subset of $\mathcal{P}(M)$.
Note that since $(\mathcal{E}^{[N]}_{\mathbf{X}^{[N]}_{t}})$ is a process
on $\mathcal{P}(M)$, the transition probabilities are elements of
$\mathcal{P}(\mathcal{P}(M))$.
  
In order to study the $N\limgo\infty$ behaviour of macroscopic variables
it is enough to study the behaviour of the process of empirical measures
in this so called {\rm thermodynamic limit}, and this is what we will do now.

\subsection{Deterministic dynamics on $\mathcal{P}(M)$ in the thermodynamic
            limit}

The first observation is that by law-of-large-number type arguments
the thermodynamic limit of the empirical
measure at some time $t+1$ can be computed from that at time $t$.

\begin{thm}
\label{th:1}
Let $(\nu^{[N]})$ be a family of probability measures in $\mathcal{P}(M)$,
and $\nu^{[N]} \stacksymb[w]{\limgo}{N\rightarrow\infty} \nu$. Then
the conditional probability 
that
\begin{equation}
\label{eq:3-20}
   \mathcal{E}^{[N]}_{\mathbf{X}^{[N]}_{t+1}}
          \stacksymb[w]{\limgo}{N\rightarrow\infty} \mathcal{F}(\nu)
\end{equation}
given $\mathcal{E}^{[M]}_{\mathbf{X}^{[M]}_{t}}=\nu^{[M]}$ for all
$M=1,2,\dots$
is equal to $1$,
where the map $\mathcal{F}:\mathcal{P}(M)\rightarrow\mathcal{P}(M)$
is defined by
\begin{equation}
\label{eq:3-30}
  \left( \mathcal{F}(\varphi) \right)(A) =
  \int_M Q_{F(x,\varphi)}(A) \varphi (\d x) .
\end{equation}
\end{thm}

Theorem \ref{th:1} means that in the thermodynamic limit the
empirical measures follow a deterministic time
evolution as described by the map $\mathcal{F}$ appearing in 
eq. (\ref{eq:3-30}).

Instead of giving a detailed proof of the theorem a few remarks will indicate
here which arguments lead to the theorem, which is a harmless generalisation
of a fundamental theorem of statistics, attributed to V.~S.~Varadarajan in
\cite{Dud}. An essential observation is that 
\begin{equation}
\label{eq:3-35}
  \Prob \{ X^{[M](j)}_{t+1}\in A| \mathcal{E}^{[M]}_{\mathbf{X}^{[M]}_{t}} 
                         = \nu^{[M]} \}
  = (\mathcal{F}(\nu^{[M]}))(A) . 
\end{equation}
It is then a consequence of the strong law of large numbers that for every
bounded function $g:M\to\Rset$ the expression
\begin{equation}
\label{eq:3-37}
  \int_{M} g(y) \mathcal{E}^{[N]}_{\mathbf{X}^{[N]}_{t+1}}(\d y)
  =
  \frac{1}{N} \sum_{j=1}^{N} g(X_{t+1}^{[N](j)})
\end{equation}
converges to
\[
  \int_{M} g(y) \mathcal{F}(\nu) (\d y)
\]
with conditional probability 1.

The last step which completes the proof of Theorem \ref{th:1} concerns the
fact that the exceptional null-sets on which the above convergence does not
hold true may depend on the function $g$. But standard topological arguments
can be used to exclude the possibility that all the different null-sets --- for
all bounded continuous functions $g$ --- do not add up to a set of positive
measure, so that the conditional probability for the weak convergence
(\ref{eq:3-20}) is 1, indeed.

In the case of globally coupled maps without noise, i.~e.~$Q_{x}=
\delta_x$, eq. (\ref{eq:3-30}) reduces to the nonlinear Frobenius
Perron equation mentioned in the Introduction, which has been
introduced by Kaneko \cite{K:D55} and by Pikovski and Kurths \cite{PKb}.
 
Many --- mostly numerical --- studies of the nonlinear Frobenius
Perron equation and its noisy counterpart show that the map
$\mathcal{F}$ often has a very complicated attractor structure
\cite{PKb,K:D55,K:D86,Mor,NKPRE}. 
Periodic, quasiperiodic, and chaotic attractors of the dynamics
(\ref{eq:3-30}) have been found and discussed in connection with
some counterintuitive observations about the temporal behaviour of
macroscopic variables. 
Also, multi-stability with the simultaneous existence of
many (possibly infinitely many) attractors is possible.

Not surprisingly it is the nonlinearity of (\ref{eq:3-30}) 
which opens up the possibility of a much richer dynamical behaviour than
what is known from the linear
Frobenius Perron equation and its noisy counterpart (which is the
limiting case of (\ref{eq:3-30}) in which $F(x,\varphi)$ depends
on x only --- and not on $\varphi$ (uncoupled case)). 

\subsection{Large deviations from the deterministic dynamics of
            empirical measures in one time step}

Theorem \ref{th:1} describes the dynamics of the empirical measures
in the thermodynamic limit deterministically by the map $\mathcal{F}$.
By definition, for large but finite $N$ the temporal evolution of 
$(\mathcal{E}^{[N]}_{\mathbf{X}^{[N]}_{t}})$ will be close to
that dynamics --- at least on a short time scale. 
It is very desirable to
make this statement more concrete and quantitative,
because then we will be in a position to derive predictions about
the behaviour of large finite systems from the properties of the
map $\mathcal{F}$. 

In the Introduction we gave an example of how the behaviour of a
macroscopic variable for a large finite system looks like a stochastically
perturbed version of its behaviour in the thermodynamic limit.
Here, we trace this observation back to the level of empirical measures:
The behaviour of $\mathcal{E}^{[N]}_{\mathbf{X}^{[N]}_{t}}$ can be understood
as the action of the map $\mathcal{F}$ plus random perturbations with
a strength that decreases with $\frac{1}{N}$. These random perturbations
have the so called {\em large deviation property}, which is essential for
the methods we will apply to study their influence.

First we recall what it means to say that probability measures have the
large deviation property \cite{DemZ,EllD}.

Consider a family of probability measures $(P^{[\varepsilon]})$ on some 
compact metric
space $\mathcal{S}$. This family is said to have the large deviation property 
with deviation rate $I$, which is a function $I:\mathcal{S}\to 
[0,\infty]$ with compact level sets $\{s\in\mathcal{S}:I(s)\leq B\}$ 
for arbitrary $B\in[0,\infty)$, if for all bounded continuous
functions $h:\mathcal{S}\to\mathbb{R}$ the following limit holds:
\begin{equation}
\label{eq:3-40}
  \lim_{\varepsilon\to 0} \varepsilon \log \int_{\mathcal{S}} 
                 \exp[-\frac{h(s)}{\varepsilon}] P^{[\varepsilon]}(\d s)
 = -\inf_{s\in\mathcal{S}}[ h(s) + I(s)] .
\end{equation}
To be precise, this property is called the {\em Laplace principle} \cite{EllD}
because
of its obvious relation to the Laplace approximation (i.e. the 
``saddle point approximation'' for real valued integrals), but (at least
in our setting) this is equivalent to the large deviation property in
its original formulation.

It is convenient and helpful for intuition to introduce the following 
symbolic notation for the essence of (\ref{eq:3-40}):
\begin{equation} 
\label{eq:3-50}
  P^{[\varepsilon]}(\d s) \asym_{\varepsilon\to 0} 
  \exp[-\frac{I(s)}{\varepsilon}] \d s .
\end{equation} 

Thus, the large deviation property can be characterised as a tool for
estimating the probability of rare events: Roughly speaking, the sets 
$A\subset\mathcal{S}$ which do not contain points where the deviation rate is
vanishing, have a probability that converges to zero exponentially fast as
$\varepsilon\to 0$, and the quantity $\inf_{s\in A}I(s)$ gives the exponential
rate of this convergence.

In the context of random perturbations of dynamical systems, a discrete-time
Markov process $S^{[\varepsilon]}_{t}$ with values in $\mathcal{S}$,
parametrised by a perturbation strength $\varepsilon$, is called a random
perturbation of a dynamical system $G:\mathcal{S}\to\mathcal{S}$ with large
deviation property, if there is a deviation rate $s\mapsto I(s|r)$
($s,r\in\mathcal{S}$) for the transition probabilities
$\Prob\{S^{[\varepsilon]}_{t+1}\in \cdot|S^{[\varepsilon]}_{t}=r\}$ which is
lower semi-continuous in $r$ and has the property that $I(s|r)=0$ iff $s=G(r)$.
The simplest example of a dynamical system with random perturbations with
large deviation property is a one-dimensional map ($\mathcal{S}=\mathbb{R}$)
with additive white Gaussian noise with variance $\varepsilon$: In this case,
the deviation rate is
\begin{equation}
\label{eq:3-60}
  I(s|r) = \frac{1}{2}|s-G(r)|^{2}  .
\end{equation}

Now we come back to the claim that
$(\mathcal{E}^{[N]}_{\mathbf{X}^{[N]}_{t}})$ is a random perturbation of
$\mathcal{F}:\mathcal{P}(M)\to\mathcal{P}(M)$ with large deviation property.
In order to specify the deviation rate for this process, which has to be a
function on $\mathcal{P}(M)\times\mathcal{P}(M)$, we use a well known concept
for measuring the dissimilarity of two measures in $\mathcal{P}(M)$, the {\em
relative entropy} or {\em Kullback-Leibler divergence} \cite{Kul}:

For $\varphi,\psi \in \mathcal{P}(M)$, the relative entropy of $\varphi$ with
respect to $\psi$ is defined as
\begin{equation}
\label{eq:3-70}
  H(\varphi|\psi) := \int_{M} \frac{\d \varphi}{\d \psi}(x) \left(
                     \log\left( \frac{\d \varphi}{\d \psi}(x) \right)
                                                            \right)
                     \psi(\d x)
\end{equation}
if $\varphi$ is absolutely continuous with respect to $\psi$, and
$H(\varphi|\psi) = \infty$ otherwise.

Note that $H(\varphi|\psi)=0$ iff $\varphi=\psi$. In information theory the
relative entropy is used as a measure for the information carried
by $\varphi$ under the hypothesis $\psi$. It can be used loosely as a way to
characterise how ``far away'' $\varphi$ is from $\psi$ in $\mathcal{P}(M)$,
although the relative entropy does not fulfil the formal definition of a
distance (as the triangle inequality does not hold even after symmetrisation).
Technically, the relative entropy is a lower semi-continuous function in its
second argument and has compact level sets as a function of its first
argument. As we will see now, for our setting with
$\mathcal{S}=\mathcal{P}(M)$ it plays the same role in the deviation rate as
the square of Euclidean distance in equation (\ref{eq:3-60}) (Gaussian noise
and $\mathcal{S}=\mathbb{R}$), which brings us to the central point of this
subsection:

\begin{thm}
\label{th:2}
The process $(\mathcal{E}^{[N]}_{\mathbf{X}^{[N]}_{t}})$ is a random
perturbation of $\mathcal{F}:\mathcal{P}(M)\to\mathcal{P}(M)$ with large
deviation property. The noise strength is $\frac{1}{N}$, and the deviation
rate:
\begin{equation}
\label{eq:3-80}
  I(\varphi|\nu) = H(\varphi|\mathcal{F}(\nu)) .
\end{equation}
\end{thm}

This theorem is a close relative of one of the classical results of
large deviation theory, the so called {\em Sanov theorem} \cite{DemZ}. 
In fact, it is
a special case of a generalised Sanov theorem proved by Dawson and
G\"{a}rtner \cite{DawG}. However, we will sketch a direct proof of Theorem
\ref{th:2}, which requires less topological prerequisites than the approach of
Dawson and G\"{a}rtner, in the Appendix.

In the symbolic notation of equation (\ref{eq:3-50}), the content of Theorem
\ref{th:2} reads like this:
\begin{equation}
\label{eq:3-90}
  \Prob\left\{\mathcal{E}^{[N]}_{\mathbf{X}^{[N]}_{t+1}} \in \d\varphi
            \left| \mathcal{E}^{[N]}_{\mathbf{X}^{[N]}_{t}} = \nu^{[N]}
       \right.\right\}
  \asym_{{N\to\infty}}
  \exp\left[-NH(\varphi|\mathcal{F}(\nu))\right] \d\varphi ,
\end{equation}
where $\displaystyle\nu=\mathrm{w-}\lim_{N\to\infty}\nu^{[N]}$.

\subsection{Large deviations from the deterministic dynamics of
            empirical measures in the long term}

Theorem \ref{th:2} gives an asymptotic estimate for the one-step
transition probability 
\[
  \Prob\left\{\mathcal{E}^{[N]}_{\mathbf{X}^{[N]}_{t+1}} \in B
            \left| \mathcal{E}^{[N]}_{\mathbf{X}^{[N]}_{t}} = \nu^{[N]}
       \right.\right\}
\] 
of the process of empirical measures 
$(\mathcal{E}^{[N]}_{\mathbf{X}^{[N]}_{t}})$ (with $B$ a Borel subset of 
$\mathcal{P}(M)$). 

The best way to understand the long term behaviour of this process is
to study invariant measures $W^{[N]}\in\mathcal{P}(\mathcal{P}(M))$ 
of the process, which are implicitly defined by
\begin{equation}
\label{eq:3-100}
  W^{[N]}(B) = \int_{\mathcal{P}(M)}
  \Prob\left\{\mathcal{E}^{[N]}_{\mathbf{X}^{[N]}_{t+1}} \in B
            \left| \mathcal{E}^{[N]}_{\mathbf{X}^{[N]}_{t}} = \mu
       \right.\right\} W^{[N]}(\d\mu)
\end{equation}
for all Borel subsets of $\mathcal{P}(M)$.

An invariant measure $W^{[N]}$ gives highest weight to those areas of
$\mathcal{P}(M)$ where the empirical measure of the system's configuration
can most likely be found in the long term.

From Theorem \ref{th:2} we know that for large $N$ the process
$(\mathcal{E}^{[N]}_{\mathbf{X}^{[N]}_{t}})$ behaves like a small random
perturbation of the deterministic map $\mathcal{F}$, and therefore we
expect that invariant measures $W^{[N]}$ are concentrated near the
attractors of $\mathcal{F}$. It is reasonable to assume that that the
family $(W^{[N]})$ itself has a large deviation property.

\begin{thm}
\label{th:3}
If a family $(W^{[N]})$ of invariant measures of the processes 
$(\mathcal{E}^{[N]}_{\mathbf{X}^{[N]}_{t}})$ has the large deviation
property with noise strength $\frac{1}{N}$ and deviation rate 
$\Phi:\mathcal{P}(M)\to [0,\infty], \varphi\mapsto\Phi(\varphi)$
(the so called {\em quasipotential}), then
\begin{equation}
\label{eq:3-110}
  \Phi(\psi) = \inf_{\varphi\in\mathcal{P}(M)} 
   \left[ H(\psi|\mathcal{F}(\varphi)) + \Phi(\varphi) \right] .
\end{equation}
\end{thm}

Formally, equation (\ref{eq:3-110}) can be interpreted as the result
of evaluating the exponential asymptotics of equation (\ref{eq:3-100}),
namely
\begin{equation}
\label{eq:3-120}
  \exp [-N\Phi(\psi)] \asym_{N\to\infty}
  \int_{\mathcal{P}(M)} \d\varphi \exp[-N H(\psi|\mathcal{F}(\varphi))]
                                  \exp[-N \Phi(\varphi)] ,
\end{equation}
by a Laplace approximation of the integral. 

Theorem \ref{th:3} can be proved like follows: 
Since the deviation
rate of a family of probability measures with large deviation property
is a unique function, it is sufficient to show that for all bounded
continuous functions $h:\mathcal{P}(M)\to\mathcal{P}(M)$ it is true
that 
\begin{eqnarray}
\label{eq:3-130}
  -\inf_{\psi\in\mathcal{P}(M)} [h(\psi)
       +\inf_{\varphi\in\mathcal{P}(M)}[H(\psi|\mathcal{F}(\varphi)
                                        +\Phi(\varphi)]] & & \nonumber\\
  = 
  \lim_{N\to\infty} \frac{1}{N}
  \log \int_{\mathcal{P}(M)} \exp[-Nh(\psi)] W^{[N]}(\d\psi) & & . 
\end{eqnarray}
But the left hand side of equation (\ref{eq:3-130}) can be written as
\[
  -\inf_{\varphi\in\mathcal{P}(M)} [\Phi(\varphi)
       +\inf_{\psi\in\mathcal{P}(M)}[H(\psi|\mathcal{F}(\varphi)
                                     +h(\psi)]] ,
\]
and an application of Theorem \ref{th:2} shows that this is equal to
\[
  -\lim_{N\to\infty} \inf_{\varphi\in\mathcal{P}(M)} \left[ \Phi(\varphi)
       - \frac{1}{N} \log \int_{\mathcal{P}(M)} \e^{-Nh(\psi)}
  \Prob\left\{\mathcal{E}^{[N]}_{\mathbf{X}^{[N]}_{t+1}} \in \d\psi
            \left| \mathcal{E}^{[N]}_{\mathbf{X}^{[N]}_{t}} = \varphi
       \right.\right\} \right] .
\]
Since $\Phi$ is the deviation rate for $(W^{[N]})$, equation (\ref{eq:3-40})
can be used to rewrite this last expression as 
\begin{eqnarray*}
  \hspace{-2cm}
  \lim_{N\to\infty} \frac{1}{N} \log
  \int_{\mathcal{P}(M)} \exp \left[ N \frac{1}{N} 
  \log \int_{\mathcal{P}(M)} \e^{-Nh(\psi)}
  \Prob\left\{\mathcal{E}^{[N]}_{\mathbf{X}^{[N]}_{t+1}} \in \d\psi
            \left| \mathcal{E}^{[N]}_{\mathbf{X}^{[N]}_{t}} = \varphi
       \right.\right\} \right] W^{[N]}(\d\varphi) & & \\
  =
  \lim_{N\to\infty} \frac{1}{N}
  \log \int_{\mathcal{P}(M)} \exp[-Nh(\psi)] W^{[N]}(\d\psi) , \hfill & &
\end{eqnarray*}
where in the last transformation we used equation (\ref{eq:3-100}).
This completes the proof of Theorem \ref{th:3}.
 
If we are able to find the quasipotential, i.e. the deviation rate 
for invariant measures of the process of empirical measures 
then we have all we need to know for exponential estimates of
the long term behaviour of empirical measures and thereby of
the long term behaviour of all global variables. We will come back to
this point in the next subsection. 

Theorem \ref{th:3}
gives the key information how to find quasipotentials:
Equation (\ref{eq:3-110}) is an eigenvalue equation in a generalised
sense, and there exists a complete theory about how to find
solutions to such equations. A very brief summary of the relevant
facts from that theory \cite{HmSem,KoM}
will be given below.

The concept of a quasipotential for an asymptotic description of the
long term behaviour of stochastic perturbations of dynamical systems
was introduced (under various names) by several authors in the context
of low-dimensional diffusion processes with a small noise term. 
Early work on quasipotentials published in the physics literature
was inspired by an analogy with semiclassical approximation in
the path integral formalism of quantum mechanics, 
and by the fact that in the weak noise limit the Fokker-Planck equation 
can be linked to a Hamilton-Jacobi equation; for references see \cite{GraR}. 
A rigorous mathematical formulation of the quasipotential method for these
continuous-time situations 
was given by Wentzell and Freidlin \cite{FW}. Later, a 
version of the quasipotential method for the discrete-time systems
was formulated \cite{Kif,HmG1}, and this is what we need here.

The usual Wentzell-Freidlin approach to quasipotentials does not start from
the {\em assumption} that there is a family of invariant measures with
the large deviation property (like Theorem \ref{th:3}) but rather 
shows its existence. However, the conditions used in the arguments 
of references \cite{Kif,HmG1} are not strictly fulfilled in the present case. 
Instead of
going into the details of these arguments again with an adjusted set
of preconditions, we pragmatically content ourselves here with
showing how to find a quasipotential if it exists, which can be
done by studying the algebraic consequences of Theorem \ref{th:3} .
The proofs of the following statements can be found in \cite{HmSem}

Define the {\em least} $n$-{\em action} (in \cite{HmSem} called the $n$-step
transition pseudodensity) between $\chi_0\in\mathcal{P}(M)$ and
$\chi_n\in\mathcal{P}(M)$ as
\begin{equation}
\label{eq:3-140}
  S_n(\chi_n|\chi_0) :=
  \inf_{\chi_1,\dots,\chi_{n-1}}
  \sum_{j=1}^{n} H(\chi_j|\mathcal{F}(\chi_{j-1}))
\end{equation}
and the {\em least action} (in \cite{HmSem} called the transitive closure of
the transition pseudodensity) between $\varphi\in\mathcal{P}(M)$ and 
$\psi\in\mathcal{P}(M)$ as
\begin{equation}
\label{eq:3-150}
  S(\psi|\varphi) :=
  \inf_{n\geq 1} S_n(\psi|\varphi) .
\end{equation}
 
The name {\em action} for a sum of one-step deviation rates like in
(\ref{eq:3-140}) alludes to the formal analogy to semi-classics and
hints at the methods which we will apply for the calculations in the next
section.

Least actions are the crucial elements in the construction of
solutions of equation (\ref{eq:3-110}):

First define the set $R_{\mathcal{F}}^{s}\subset\mathcal{P}(M)$ as 
the set of measures $\psi$ which have the following properties:
\begin{enumerate} 
 \item $S(\psi|\psi)=0$.
 \item If any measure $\varphi\in\mathcal{P}(M)$ fulfils 
       $S(\varphi|\psi)=0$, it is also true that $S(\psi|\varphi)=0$.
\end{enumerate} 
The set $R_{\mathcal{F}}^{s}$ is related to and in many cases identical
to the union of attractors of the map $\mathcal{F}$; for details about
this relation see \cite{HmSem}.

Every solution of (\ref{eq:3-110}) can be written in the following
form:
\begin{equation}
\label{eq:3-160}
  \Phi(\varphi) = 
  \inf_{\alpha\in R_{\mathcal{F}}^{s}} 
  \left[ S(\varphi|\alpha) + c(\alpha) \right] ,
\end{equation}
where the coefficients $c(\alpha)$ are free parameters, if one is only
interested in solving (\ref{eq:3-110}). But since the quasipotential
is the deviation rate for the invariant measures, the coefficients have
to fulfil some additional equations that guarantee compatibility 
with (\ref{eq:3-100}), and these equations fix the values $c(\alpha)$ 
up to an additive constant which can be set to zero. A graph theoretical
method of how to determine the values $c(\alpha)$ when the least
actions $V(\psi|\alpha)$ are known for all $\alpha\in R_{\mathcal{F}}^{s}$
goes back to Wentzell and Freidlin \cite{FW}. We will not explain the details
of this procedure, for which we refer to \cite{FW,Kif,HmG1}, as in a general 
context it is enough to know that the coefficients can be determined in
principle, while the
concrete example which we will study at the end of the next section, has a map
$\mathcal{F}$ for which the only attractor is a stable fixed point
$\alpha^*$. In this case we have obviously:
\begin{equation}
\label{eq:3-170}
  \Phi(\psi) = S(\psi|\alpha^*) .
\end{equation}  

\subsection{Large deviations from the deterministic dynamics of
            global variables in the long term}

The quasipotential $\Phi$ describes the long term behaviour of
empirical measures as $N$ approaches infinity, or more precisely
the probability for finding certain empirical measures in the
long time development of the system. From a practical point of view,
however, it is cumbersome to observe probability distributions 
on the infinite dimensional space $\mathcal{P}(M)$ in large coupled 
systems.  Instead, it is usually more convenient
to follow the time development of (low dimensional) global variables,
i.e. variables that depend on the empirical measure through a
continuous  map
\begin{equation}
\label{eq:3-180}
  G: \mathcal{P}(M) \to \Rset^{k} .
\end{equation} 
We have mentioned already two examples: mean value (\ref{eq:2-70}) and
mean field (\ref{eq:2-90}).

The invariant measure $W^{[N]}$ of the process of empirical measures fixes 
a measure $W_{G}^{[N]}$ on $G(\mathcal{P}(M))\subset \Rset^{k}$:
\begin{equation}
\label{eq:3-185}
  W_{G}^{[N]}(\hat{B})
  = W^{[N]}(\{\varphi\in\mathcal{P}(M): G(\varphi)\in\hat{B}\})
\end{equation}
for all Borel subsets $\hat{B}$ of $G(\mathcal{P}(M))$. $W_{G}^{[N]}$
describes the long term behaviour of the global variable $G$.

The fact that $(W^{[N]})$ has the large deviation property implies a
large deviation property for $(W_{G}^{[N]})$, too, and the deviation
rate can be found through a theorem which is known as {\em contraction
principle} in large deviation theory \cite{DemZ,EllD}:

\begin{thm}
\label{th:4}
The deviation rate of the family $(W_{G}^{[N]})$ of probability measures
describing the long term statistics of the global variable $G$, 
which we call {\em contracted quasipotential} $\phi_{G}$,
can be derived in the following way from the quasipotential $\Phi$:
\begin{equation}
\label{eq:3-190}
  \phi_{G}(u) = \inf_{\psi\in\mathcal{P}(M): G(\psi)=u } 
  \left[ \Phi(\psi) \right]
\end{equation}
for all $u\in G(\mathcal{P}(M))$.
\end{thm}

The proof of this theorem consists in the simple observation that 
the large deviation property (\ref{eq:3-40}) for $(W^{[N]})$ implies that
for all bounded continuous functions $h_{G}:G(\mathcal{P}(M))\to\Rset$
one has:
\begin{eqnarray}
\label{eq:3-200}
  \lim_{N\to\infty} \frac{1}{N} &\log& \int_{\mathcal{P}(M)}
                 \exp[-N h_{G}(G(\psi))] W^{[N]}(\d \psi) 
  \nonumber \\
 & = & -\inf_{\psi\in\mathcal{P}(M)}[ h_{G}(G(\psi)) + \Phi(\psi)] .
\end{eqnarray} 
But the left hand side of equation (\ref{eq:3-200}) is equal to
\[
  \lim_{N\to\infty} \frac{1}{N} \log \int_{G(\mathcal{P}(M))}
                 \exp[-N h_{G}(u)] W_{G}^{[N]}(\d u)
\]
whereas the right hand side is equal to
\[
 -\inf_{u\in G(\mathcal{P}(M))}[ h_{G}(u) + \inf_{G(\psi)=u}\Phi(\psi)]
\]
and this shows that the statement of Theorem \ref{th:4} is true.

Sometimes it is possible to recover part of the dynamical information lost 
during the `projection' of elements of the infinite dimensional space
$\mathcal{P}(M)$ to $k-$dimensional values of the global variable by
studying a {\em time delay} plot as mentioned in the Introduction.
For instance, a two dimensional time delay plot of a one dimensional
global variable $G$ would mean that for some realization of the
process $(\mathbf{X}^{[N]}_{t})$ one plots the points
\[
  \left( G(\mathcal{E}^{[N]}_{\mathbf{X}^{[N]}_{t}}),
         G(\mathcal{E}^{[N]}_{\mathbf{X}^{[N]}_{t+1}}) \right)
\]
in a two dimensional plane.

In the long term, the time delay plot of $G$ is expected to show 
the set 
\begin{equation}
\label{eq:3-210}
  \{(u,v)\in\Rset\times\Rset | u=G(\varphi),v=G(\mathcal{F}(\varphi)),
                               \varphi\in R_{\mathcal{F},1}^{s} \}
\end{equation}
in the thermodynamic limit, where $R_{\mathcal{F},1}^{s}$ is one of 
the attractors of $\mathcal{F}$.

For finite $N$, however, the points in the time delay plot will be
distributed according to a probability measure $W_{G,2}^{[N]}$ on 
$\Rset\times\Rset$. Again, the measures $(W_{G,2}^{[N]})$ have the
large deviation property, and it follows from Theorem \ref{th:2},
Theorem \ref{th:3}, and equation (\ref{eq:3-40}) 
that the corresponding deviation rate, the {\em time delay contracted
quasipotential $\phi_{G,2}$}, can be calculated from the quasipotential
$\Phi$ in the following way: 
\begin{equation}
\label{eq:3-220}
  \phi_{G,2}(u,v) = \inf_{G(\varphi)=u,G(\psi)=v}
                    [\Phi(\varphi) + H(\psi|\mathcal{F}(\varphi))] .
\end{equation}
This time delay contracted quasipotential predicts the outcome of
time delay plots for large but finite $N$. Its minimum or minima coincide
with the set (\ref{eq:3-210}), and its level cuts (contour lines) in heights
proportional to $1/N$ determine the regions of the plane in which points
of the time delay plot are to be expected when $N$ is finite. 

In the example of Fig.~1 , the time delay contracted quasipotential is
expected to have a ring-shaped valley along the thermodynamic limit attractor
and a hill in the middle of that ring, the height of which is proportional to
the critical number of subsystems which is necessary in order to resolve the
ring structure in a simulation.

\section{Approximative computation of contracted quasipotentials}
\label{se:4}

In the preceding section we have built the theoretical fundament for
a large deviation type description of the $N\to\infty$ asymptotics for
the long term behaviour of globally coupled maps. The central 
object in this description is the quasipotential; however, its
computation is a complex and difficult minimisation problem over paths
(i.e., sequences) in the infinite dimensional space $\mathcal{P}(M)$ of
probability measures. This is a much more difficult task than the
minimisation problem that one has to solve in order to compute
quasipotentials for low dimensional maps. Nevertheless, a Hamiltonian
approach to the minimisation problem, which has proved to be 
successful in the low-dimensional context, can be useful for
the present situation, too. However, in order to reach at
practical algorithms we do not simply copy the low dimensional
procedure to treat the too complicated 
infinite-dimensional problem of computing
$\Phi(\varphi)$, but we describe a way to approximate computations
of contracted quasipotentials $\phi_{G}(u)$, which are defined on a 
low-dimensional space.  

\subsection{A Hamiltonian formalism for minimising actions}

For the following we assume that the map $\mathcal{F}$ has only one
attractor and that $\chi_0$ is a point of this attractor. This 
simplifies the notation, and the general case of more than one
attractor is complicated only by the fact that several pieces of
`local' quasipotentials have to be stitched together like in
equation (\ref{eq:3-160}).

We concentrate now on the problem how to compute the contracted
quasipotential $\phi_{G}$ for a global variable $G$.
According to equations (\ref{eq:3-190}), (\ref{eq:3-170}), (\ref{eq:3-150}),
and (\ref{eq:3-140}) we have the following multiple minimisation
problem:
\begin{equation}
\label{eq:4-10}
  \phi_{G}(u) =
  \ \inf_{n\geq 1} \ \inf_{G(\chi_{n})=u}
  \ \inf_{\chi_{1},\dots,\chi_{n-1}}
  \  \sum_{j=1}^{n} H(\chi_{j}|\mathcal{F}(\chi_{j-1})) ,
\end{equation} 
where all $\chi_{j}$ are elements of $\mathcal{P}(M)$.

It will turn out to be useful to introduce in an intermediate step
constraints not only to the value of the global variable in the
last state $\chi_{n}$ but in all steps of the minimising sequence
$(\chi_{j})$:. Defining for $u_{1},\dots,u_{n}\in \Rset^{k}$
the constrained $n-$action
\begin{equation}
\label{eq:4-20}
  S_{G,n}(u_{1},\dots,u_{n}|\chi_{0}) :=
  \ \inf_{G(\chi_{1})=u_{1},\dots,G(\chi_{n})=u_{n}}
   \ \sum_{j=1}^{n} H(\chi_{j}|\mathcal{F}(\chi_{j-1})) ,
\end{equation}
we can rewrite equation (\ref{eq:4-10}) as
\begin{equation}
\label{eq:4-30}
  \phi_{G}(u) =
  \ \inf_{n\geq 1} \ \inf_{u_{1},\dots,u_{n-1}}\
   S_{G,n}(u_{1},\dots,u_{n-1},u|\chi_{0}) .
\end{equation}
The advantage of the introduction of constrained $n-$actions is that
the minimisation problem of equation (\ref{eq:4-30}) looks very much
like the problem of determining the quasipotentials of a low-dimensional
(i.e., $k-$dimensional) map, the only difference being that the
$n-$action of a low-dimensional map has been replaced by a constrained
$n-$action derived from an infinite dimensional map.

This shows that the main new technical problem which we face is
the computation of constrained $n-$actions (\ref{eq:4-20}), while 
the step (\ref{eq:4-30}) can then be solved by standard methods. 

Equation (\ref{eq:4-20}) which defines the constrained $n-$actions
can be reformulated as a Lagrangian minimisation problem; we
do this here under the assumption that $M$ is a compact interval
in $\Rset$ and that the probability measure $\chi_{0}$ is absolutely
continuous with respect to the Lebesgue measure on $M$ so that
$\chi_{0}$ has a probability density $\rho_{0}$.  

In this case we can write
\begin{equation}
\label{eq:4-40}
  S_{G,n}(u_{1},\dots,u_{n}|\chi_{0})
  = \ \min_{\rho_{1},\dots,\rho_{n}}\
  \sum_{j=1}^{n-1} \mathcal{L}_{u_{j},\lambda_{j},\mu_{j}}
      [\rho_{j-1},\rho_{j}-\rho_{j-1}]
\end{equation}
where $\rho_{1},\dots,\rho_{n}$ are integrable functions, and
a Lagrange functional
\begin{eqnarray}
\label{eq:4-50}
  \mathcal{L}_{u,\lambda,\mu}[\rho,\nu] & := &
  \int_M \{\rho(x)+\nu(x)\} \log 
             \frac{\rho(x)+\nu(x)}{(\mathcal{F}[\rho])(x)} \d x \\
  \nonumber
  & + & (\lambda - 1) \left[ \int_M \{\rho(x)+\nu(x)\} \d x - 1 \right] 
  +  \mu \left( G[\rho+\nu] - u \right)
\end{eqnarray} 
has been introduced which contains Lagrange parameters $(\lambda-1)$ and
$\mu$. As usual all Lagrange parameters $\lambda_{j}$ and $\mu_{j}$ will 
finally be fixed so that the constraints
\begin{equation}
\label{eq:4-60}
  \int_M \rho_{j}(x) \d x = 1
\end{equation}
and
\begin{equation}
\label{eq:4-70}
  G[\rho_{j}] = u_{j}
\end{equation}
are fulfilled. Note that in equations (\ref{eq:4-50}) and (\ref{eq:4-70})
the symbols $G$ and $\mathcal{F}$, which originally denoted
maps on $\mathcal{P}(M)$, are now used for the corresponding functionals
on the space of densities. 

In order to find a minimising sequence of densities $\rho_{1},\dots,
\rho_{n}$ that solves the minimisation problem (\ref{eq:4-40}) one
could start to write down the variational derivatives 
$\frac{\delta}{\delta\rho_{i}(y)}$ of the sum in (\ref{eq:4-40})
which have to be equal to zero for the minimising sequence. In this
way one would be led to an implicit relation between $\rho_{j+1}$,
$\rho_{j}$, and $\rho_{j-1}$, which is a discrete-time version of
a Lagrange equation. However, this relation cannot simply be turned
into an explicit iterative rule how to proceed from $\rho_{j-1}$ and
$\rho_{j}$ to $\rho_{j+1}$ in a minimising sequence.

The situation is easier to analyse in a Hamiltonian formulation
which can be derived from the Lagrangian one if the global variable
has the following structure
\begin{equation}
\label{eq:4-80}
  G[\rho] = \int_M \gamma(x)\rho(x) \d x ,
\end{equation}
i.e., if it is obtained by averaging some integrable function $\gamma(x)$.
This is the case for the examples: mean value (\ref{eq:2-70}) with
$\gamma(x)=x$, and mean field (\ref{eq:2-90}) with $\gamma(x)=f(x)$. 

By Legendre transforming (\ref{eq:4-50}) one arrives at the Hamiltonian
\begin{eqnarray}
\label{eq:4-90}
  \mathcal{H}_{u,\lambda,\mu}[\rho,\pi] & = &
    \int_M \left( \mathcal{F}[\rho] \right)(y) 
                 e^{\pi(y)-\lambda-\mu\gamma(y)} \d y \\
  \nonumber
  & - & \int_M \pi(y)\rho(y)\d y + (\lambda-1) + \mu u .
\end{eqnarray}  
Here, a new `momentum field' $\pi(x)$ has been introduced, so that
now the minimising sequence $(\rho_{i})$ is searched for together
with a sequence $(\pi_{i})$. This will be done by using discrete time
Hamilton equations:
\begin{eqnarray}
\label{eq:4-100}
  \rho_{j+1}(x)-\rho_{j}(x) & = &
  \left. \frac{\delta}{\delta\pi(x)} 
  \mathcal{H}_{u_{j+1},\lambda_{j+1},\mu_{j+1}}[\rho,\pi]
  \right|_{\rho=\rho_{j},\pi=\pi_{j+1}} ,\\
\label{eq:4-110}
  \pi_{j+1}(x)-\pi_{j}(x) & = &
  \left. -\frac{\delta}{\delta\rho(x)}
  \mathcal{H}_{u_{j+1},\lambda_{j+1},\mu_{j+1}}[\rho,\pi]
  \right|_{\rho=\rho_{j},\pi=\pi_{j+1}} ,
\end{eqnarray}
which leads to:
\begin{eqnarray}
\label{eq:4-120}
  \rho_{j+1}(x) & = &
  \mathcal{F}[\rho_{j}](x)\,
  e^{\pi_{j+1}(x)-\lambda_{j+1}-\mu_{j+1}\gamma(x)} , \\
\label{eq:4-130}
  \pi_{j}(x) & = &
  \int_{M} e^{\pi_{j+1}(y)-\lambda_{j+1}-\mu_{j+1}\gamma(y)}
  \frac{\delta\mathcal{F}[\rho_{j}](y)}{\delta\rho_{j}(x)} \d y .
\end{eqnarray}
Note, that a deterministic orbit of densities $\rho_0,\mathcal{F}[\rho_0],
\mathcal{F}[\mathcal{F}[\rho_0]],\dots$, which is a trivial
minimising sequence, is a solution of
equations (\ref{eq:4-120}) and (\ref{eq:4-130}) with $\pi_{j+1}=1$,
$\lambda_{j+1}=1$, and $\mu_{j+1}=0$ for all $j$.

Still, equations (\ref{eq:4-120}) and (\ref{eq:4-130}) cannot be used
for an explicit forward iteration which would generate nontrivial minimising
sequences $(\rho_{j})$ and $(\pi_{j})$, since (\ref{eq:4-130}) cannot
be inverted to give $\pi_{j+1}$ when knowing $\pi_{j}$ and $\rho_{j}$.

Nevertheless, equation (\ref{eq:4-120}) reveals important insight into
the structure of the problem. First of all, this equation shows that
the one-step action $H(\rho_{j+1}|\mathcal{F}[\rho_{j}])$
has the following form:
\begin{equation}
\label{eq:4-140}
  H(\rho_{j+1}|\mathcal{F}[\rho_{j}])
  = \int_M \rho_{j+1}(x) \pi_{j+1}(x) \d x - \lambda_{j+1}
  - \mu_{j+1} u_{j+1}  .
\end{equation}

Secondly, we can use (\ref{eq:4-120}) in the constraints (\ref{eq:4-60}) and
(\ref{eq:4-70}) in order to
see that the Lagrange parameters are determined by the following function:
\begin{equation}
\label{eq:4-150}
  \mathcal{Z}_{\pi}(\rho,\mu)
  := \int_M \mathcal{F}[\rho](x) e^{(\pi(x) - \mu\gamma(x))} \d x :
\end{equation}
$\mu_{j+1}$ is implicitly given by
\begin{equation}
\label{eq:4-160}
  u_{j+1} = - \left.\frac{\partial}{\partial\mu} 
              \log \mathcal{Z}_{\pi_{j+1}}(\rho_{j},\mu)
              \right|_{\mu=\mu_{j+1}}
\end{equation}
while
\begin{equation}
\label{eq:4-170}
  \lambda_{j+1} = \log \mathcal{Z}_{\pi_{j+1}}(\rho_{j},\mu_{j+1}) .
\end{equation}

We note in passing that equation (\ref{eq:4-130}), too, can be expressed
in terms of $\log \mathcal{Z}_{\pi_{i}}(\rho,\mu)$
\begin{equation}
\label{eq:4-180}
  \pi_{j}(x) = \left.\frac{\delta}{\delta\rho(x)}
               \log \mathcal{Z}_{\pi_{j+1}}(\rho,\mu_{j+1})
               \right|_{\rho=\rho_{j}} ,
\end{equation}
and that equations (\ref{eq:4-160}), (\ref{eq:4-170}), and (\ref{eq:4-180})
in equation (\ref{eq:4-140}) lead to an expression for the
constrained $n$-action (\ref{eq:4-20}) which involves the Legendre transform
$\mathcal{G}_{\pi_{i}}(\pi,u)$ of the function 
$\log \mathcal{Z}_{\pi_{i}}(\rho,\mu)$:
\begin{eqnarray}
  \nonumber
  S_{G,n}(u_{1},\dots,u_{n}|\chi_{0})
  & = &
  \sum_{j=1}^{n-1} \mathcal{G}_{\pi_{j+1}}(\pi_{j},u_{j+1})  \\ & &
  - \log \mathcal{Z}_{\pi_{1}}(\rho_{0},\mu_{1})
  + \mu_{1} u_{1} + \int_{M} \rho_{n}(x) \pi_{n}(x) \d x .
\label{eq:4-190}
\end{eqnarray} 
This remark shows that the present formalism can be interpreted as a
generalisation of the classic Cram\'{e}r's theorem which says that the
deviation rate of the probability laws of mean values of independent
identically distributed random variables
is the Legendre transform of the logarithmic moment generating function
of those random variables. However, on the level of practical calculations,
expression (\ref{eq:4-190}) does not offer any additional help.

For any more explicit progress in evaluating the constrained $n$-action
(\ref{eq:4-20}) it is necessary to resort to some approximations. First of
all we expand $\mathcal{Z}_{\pi_{j+1}}(\rho_{j},\mu)$ in powers of $\mu$. 
Retaining only terms up to second order in $\mu$, we are led to
approximate expressions for $\lambda_{j+1}$ and $\mu_{j+1}$. With the
abbreviations
\begin{eqnarray}
\label{eq:4-200a}
  \tilde{n}'_{j} & := & \int_{M} \mathcal{F}[\rho_{j}](x)
                 e^{\pi_{j+1}(x)} \d x ,
\\
\label{eq:4-200b}
  \tilde{u}'_{j} & := & \frac{1}{\tilde{n}'_{j}}
                 \int_{M} \gamma(x)\mathcal{F}[\rho_{j}](x)
                 e^{\pi_{j+1}(x)} \d x ,
\\
\label{eq:4-200c}
  \tilde{v}'_{j} & := & \frac{1}{\tilde{n}'_{j}}
                 \int_{M} (\gamma(x))^{2}\mathcal{F}[\rho_{j}](x)
                 e^{\pi_{j+1}(x)} \d x  - (\tilde{u}'_{j})^{2} ,
\end{eqnarray}
we obtain
\begin{equation}
\label{eq:4-210}
  \log \mathcal{Z}_{\pi_{j+1}}(\rho_{j},\mu) \approx
   \log \tilde{n}'_{j} - \mu \tilde{u}'_{j} 
       + \frac{1}{2} \mu^{2} \tilde{v}'_{j} .
\end{equation}
From equation (\ref{eq:4-160}) we see now that
\begin{equation}
\label{eq:4-220}
  \mu_{j+1} \approx -\frac{u_{j+1}-\tilde{u}'_{j}}{\tilde{v}'_{j}} ,
\end{equation}
which means that the present approximation makes sense as long as
$u_{j+1}$ stays close to $\tilde{u}'_{j}$,
and from equation (\ref{eq:4-170}) that
\begin{equation}
\label{eq:4-230}
  \lambda_{j+1} \approx \log \tilde{n}'_{j}
    + \frac{u_{j+1}-\tilde{u}'_{j}}{\tilde{v}'_{j}} 
      \left( \frac{1}{2} u_{j+1} + \frac{1}{2} \tilde{u}'_{j} 
      \right) .
\end{equation}
The corresponding approximation for the one-step action (\ref{eq:4-140}) is
\begin{equation}
\label{eq:4-240}
  H(\rho_{j+1}|\mathcal{F}[\rho_{j}]) \approx
    \int_M \rho_{j+1}(x) \pi_{j+1}(x) \d x - \log \tilde{n}'_{j}
    + \frac{|u_{j+1}-\tilde{u}'_{j}|^{2}}{2 \tilde{v}'_{j}} .
\end{equation}

At this point it is useful to look back to equation (\ref{eq:4-120})
and to note that there are two ways in which 
$\rho_{j+1}$ deviates from $\mathcal{F}[\rho_{j}]$: If the exponential
factor on the right hand side of the equation would not contain an 
$x$-dependent $\pi_{j+1}(x)$ part, $\rho_{j+1}$
would be among all densities with the prescribed value $u_{j+1}$ for the global
variable $G$ that one which has smallest relative entropy with respect to
$\mathcal{F}[\rho_{j}]$. The additional effect of $\pi_{j+1}(x)$ is to allow
for the fact that for minimising the total $n$-step action it is usually 
advantageous not to try to minimise the one-step actions 
$H(\rho_{j+1}|\mathcal{F}[\rho_{j}])$ in all the single steps but to look
at the overall effect of the choice of $\rho_{j+1}$
on the total $n$-step action.

However, it is the non-trivial influence of $\pi_{j+1}(x)$ that makes it 
so difficult to find a minimising sequence which fulfils equations 
(\ref{eq:4-120}) and (\ref{eq:4-130}). The situation becomes much simpler
if we neglect the influence of $\pi_{j+1}(x)$ by just concentrating on
eq.~(\ref{eq:4-120}) and setting
\begin{equation}
\label{eq:4-245} 
  \pi_{j}(x) \equiv 1 \mbox{ for all } j 
\end{equation}
in this equation.
This is expected to lead to a reasonable approximation of the contracted 
quasipotential in the vicinity of the attractor in the dynamics of the
global variable $G$, since in this case a minimising sequence of densities
$\rho_{j}$ should stay close to the attractor so that there is not much
scope for reducing the action by changing the sequence through the influence
of $\pi_{j}$. In any case, the approximate contracted quasipotential computed
with the assumption (\ref{eq:4-245}) is an upper bound for the true contracted
quasipotential.  

With (\ref{eq:4-245}) we obtain from equations (\ref{eq:4-200a}) --- 
(\ref{eq:4-200c}):
\begin{eqnarray}
\label{eq:4-250}
  \tilde{n}'_{j} & \approx & 1 ,
\\
\label{eq:4-260}
  \tilde{u}'_{j} & \approx & u'_{j} =
                 \int_{M} \gamma(x)\mathcal{F}[\rho_{j}](x) \d x ,
\\
\label{eq:4-270}
  \tilde{v}'_{j} & \approx & v'_{j} =
                 \int_{M} (\gamma(x))^{2}\mathcal{F}[\rho_{j}](x) \d x  
                          - (u'_{j})^{2} .
\end{eqnarray}
From (\ref{eq:4-240}) we see that the one-step action along such a 
minimising sequence can be approximated as
\begin{equation}
\label{eq:4-280}
  H(\rho_{j+1}|\mathcal{F}[\rho_{j}]) \approx
    \frac{|u_{j+1}-u'_{j}|^{2}}{2 v'_{j}} .
\end{equation}

Using this approximation for calculating the constrained $n-$action
(\ref{eq:4-20}), we can go back to equation (\ref{eq:4-30}) and obtain
\begin{equation}
\label{eq:4-290}
  \phi_{G}(u) \approx
  \ \inf_{n\geq 1} \ \inf_{u_{1},\dots,u_{n-1}}\
  \sum_{j=1}^{n} \frac{|u_{j}-u'_{j-1}|^{2}}{2 v'_{j-1}} .
\end{equation}
We have written this equation in a form which resembles
the well studied situation of a quasipotential for the weak noise
asymptotics of a low dimensional system with white Gaussian noise 
($x_{t+1}=F(x_{t})+\xi$) \cite{RT91,HmG1}. 
In that case, a quasipotential $\Phi(x)$ which
describes the asymptotic behaviour of the system's invariant density can be
computed as follows: 
\begin{equation}
\label{eq:4-300}
  \Phi(x) =
  \ \inf_{n\geq 1} \ \inf_{q_{1},\dots,q_{n-1}}\
  \sum_{j=1}^{n} \frac{|q_{j}-F(q_{j-1})|^{2}}{2} , 
\end{equation}
where in the sum on the right hand side $q_{0}$ is a point of the attractor 
of $F$, and $q_{n}=x$.

Despite this similarity, there is a crucial difference between equation
(\ref{eq:4-290}) and equation (\ref{eq:4-300}): The points $u'_{j}$ which
appear in equation (\ref{eq:4-290}) are not simply images of the points
$u_{j}$ under some low dimensional map $F$. Instead, in order to calculate
$u'_{j}$ one has to know the full density $\rho_{j}$. Therefore it is not
enough to look only at the sequence of points $(u_{j})$ but one has to follow
the sequence of densities $(\rho_{j})$, which, however, is determined by
equation (\ref{eq:4-120}) with (\ref{eq:4-245}). Unfortunately this
means that in order to evaluate (\ref{eq:4-290}) one cannot simply use the
known algorithms for computing quasipotentials according to
(\ref{eq:4-300}). Nevertheless, solving equation (\ref{eq:4-290}) is much
easier than solving the original problem (\ref{eq:4-10}), since the
minimisation now requires only variation of a sequence of points instead of
variation of a sequence of densities.

With equation (\ref{eq:4-290}) we have found a generally applicable
procedure how to compute an upper bound for contracted quasipotentials
which can serve as a good estimate for the exact contracted quasipotential
near attractors.

\subsection{A simple example}

In this subsection we study a very simple model system of globally coupled
noisy maps. The purpose of this example is, on the one hand, to illustrate the 
definitions and computational procedures that were introduced in this article,
and, on the other hand, to serve as a test for the quality of the
approximation formula (\ref{eq:4-290}).

The model system consists simply of linear maps on $\Rset$:
\begin{equation}
\label{eq:4b-10}
  f_{a}(x) = ax, \ \ \ -1<a<1,
\end{equation}
to which white Gaussian noise with a standard deviation $\sigma$ is added, 
and which are coupled through a
function $k_{\kappa}:\Rset\to\Rset$ of their mean field (where $\kappa$ is a
coupling strength and $k_{0}(h)=0$ for all $h\in\Rset$); this means that
equation (\ref{eq:2-30}) has to be applied with the function 
\begin{equation}
\label{eq:4b-20}
  F_{\kappa}(x,\mu) = f_{a}(x) + k_{\kappa}(\tilde{h}(\mu))
                    = ax +  k_{\kappa}(\tilde{h}(\mu))
\end{equation}
(cf. (\ref{eq:2-90})), and equation (\ref{eq:3-30}) becomes:
\begin{equation}
\label{eq:4b-30}
  \mathcal{F}[\rho](y) = \frac{1}{\sqrt{2\pi}\sigma}
                         \int_{\Rset} 
                         e^{-\frac{1}{2\sigma^{2}}
                            (y-ax-k_{\kappa}(\tilde{h}[\rho]))^{2}}
                         \rho(x) \d x .
\end{equation}
The subsystem state space of this example is $\Rset$ rather than a
compact space --- in contrast to an assumption that was used
for the derivations in the previous sections; the reason for choosing such a
system is the simplicity of calculations which serves the purpose of
illustrating the various steps in the construction of quasipotentials.
Therefore we do not bother here to give proofs for the non-compact case. In
practise it is not difficult (for sufficiently small coupling strength) to
modify the system in a way that the subsystems are restricted to a closed
interval by cutting of the noise whenever a subsystem would leave that
interval. Such a modification does not change the asymptotic behaviour
and therefore it is convenient to use the original system for calculations.  

We denote a Gaussian probability density with mean value $m$ and variance $v$
by 
\begin{equation}
\label{eq:4b-40}
  \mathcal{N}_{m,v}(y) := \frac{1}{\sqrt{2\pi v}}
                          e^{-\frac{1}{2v}(y-m)^{2}} .
\end{equation}
Now it is an important observation that Gaussian densities stay Gaussian under
the influence of the temporal evolution (\ref{eq:4b-30}). In fact, we have
\begin{equation}
\label{eq:4b-50}
  \mathcal{F}[\mathcal{N}_{m,v}]  =  \mathcal{N}_{\hat{m},\hat{v}}
\end{equation}
with
\begin{eqnarray}
  \widehat{m} & = & a m + k_{\kappa}(a m) ,
\label{eq:4b-60} \\
  \widehat{v} & = & a^{2} v + \sigma^{2} .
\label{eq:4b-70}
\end{eqnarray}
Equation (\ref{eq:4b-70}) has a stable fixed point
\begin{equation}
\label{eq:4b-80}
  v^{*} = \frac{\sigma^{2}}{1-a^{2}}.
\end{equation}
If $m^{*}$ is any point of an attractor of (\ref{eq:4b-60}),
the density $\mathcal{N}_{m^{*},v^{*}}$ is a point of an attractor
of $\mathcal{F}$. For a linear coupling, $k_{\kappa}(h)=\kappa h$,
one has $m^{*}=0$, but for nonlinear couplings there may be different
attractors of (\ref{eq:4b-60}).

We now want to find the contracted quasipotential $\phi_{\tilde{m}}$ for the
global variable $\tilde{m}$, the mean value, i. e. $\gamma(x)=x$ in equation
(\ref{eq:4-80}). Starting from the density
$\rho_{0}=\mathcal{N}_{m^{*},v^{*}}$, one has to look for a minimising
sequence of densities $\rho_{j}$ according to equation (\ref{eq:4-120}) and
fulfilling the constraints (\ref{eq:4-70}) for a given sequence of mean values
$(u_{j})$.

If we use the simplification (\ref{eq:4-245}), the minimising sequence turns
out to be the sequence of Gaussian densities $\mathcal{N}_{u_{j},v^{*}}$. Then,
the approximation (\ref{eq:4-290}) for the contracted quasipotential reads
like this:
\begin{equation}
\label{eq:4b-90}
  \phi_{\tilde{m}}(m) \approx
  \ \inf_{n\geq 1} \ \inf_{u_{1},\dots,u_{n-1}}\
  \frac{1}{2v^{*}}\sum_{j=1}^{n} |u_{j}-\widehat{u_{j-1}}|^{2} , 
\end{equation}
with $u_{0}=m^{*}$, $u_{n}=m$, where the connection between $u_{j}$ and
$\widehat{u_{j}}$ is like in equation (\ref{eq:4b-60}).

This means that in this case the approximate contracted quasipotential is
--- up to the constant factor $\frac{1}{v^{*}}$ --- equal to the
quasipotential of the one-dimensional map $m\mapsto am + k_{\kappa}(am)$ with
Gaussian random perturbations, which can be calculated using the well
established methods of \cite{RT91} and \cite{HmG1}.   

In the case of linear coupling, $k_{\kappa}(h)=\kappa h$, this map is linear:
$m\mapsto (1+\kappa)am$, and equation (\ref{eq:4b-90}) can be solved
explicitely: 
\begin{equation}
\label{eq:4b-100}
  \phi_{\tilde{m}}(m) \approx
  \frac{1-(1+\kappa)^{2}a^{2}}{2v^{*}} m^{2}.
\end{equation} 

This result can be checked through numerical simulations of the system of
coupled maps (\ref{eq:2-30}) with (\ref{eq:4b-20}): The data of a long
time simulation of the $N$ subsystem can be used for plotting a histogram
$\hat{w}_{\tilde{m}}(m)$ of the distribution of mean values, and Theorem
\ref{th:4} tells us that $-\frac{1}{N}\log\hat{w}_{\tilde{m}}(m)$ should 
converge
to the contracted quasipotential as $N\to\infty$. Typically, one expects that
already for large finite $N$ the histogram
$-\frac{1}{N}\log\hat{w}_{\tilde{m}}(m)$ shows the shape of the quasipotential,
shifted by an additive offset due to the pre-factor hidden in asymptotic
relations like (\ref{eq:3-50}), which is only weakly dependent on $m$.

Fig.~3 compares the approximation (\ref{eq:4b-100}) for
the contracted quasipotential with histograms obtained in simulations. Part a)
shows the case $\kappa=0$, $a=0$ of uncoupled identically distributed random
variables, where the result $\phi_{\tilde{m}}(m)=\frac{1}{2\sigma^{2}}m^{2}$
is not only a special case of our approximation (\ref{eq:4b-100}) but a
necessary consequence  of the central limit theorem. The histogram contains
the results of simulations with $N=10^{5}$ and $\sigma=1$, and its shape fits
very well with the shifted contracted quasipotential, which is the solid line
parabola in the figure.  

Part b) is for the less trivial case $\kappa=0.4$, $a=0.5$. Here and in all
the following examples we use again $\sigma=1$. The lower half of
the figure shows a histogram for a simulation with $N=10^{5}$, the upper half
with $N=10^{6}$. While the parabolic contracted quasipotential
(\ref{eq:4b-100}) gives a reasonable approximation to the histogram for
$N=10^{5}$, it obviously works better for $N=10^{6}$.

Looking at nonlinear coupling functions $k_{\kappa}(h)$ one can study
more complicated mean field dynamics. With the choice
\begin{equation}
\label{eq:4b-110}
  k_{\kappa}(h)= 10\kappa (1 - h - \alpha h^{2})
\end{equation}
with some nonlinearity parameter $\alpha$ one has for $\kappa = 0.1$ the
following mean field map: $m\mapsto 1-\alpha a^{2} m^{2}$. This is the
logistic map, and so the temporal evolution of the mean field can show all
the dynamical behaviours that are known for the logistic map. The
approximation (\ref{eq:4b-90}) for the contracted quasipotential is the
conventional quasipotential for the logistic map, which cannot be written down
explicitely but can be solved numerically with high precision. 

Fig.~4 shows the results for this nonlinearly coupled system of noisy
linear maps with $a=0.5$, $\kappa=0.1$ and $\alpha=3.16$. For this set of
parameters the mean field map has a stable period 2. Part a) of the figure
shows the approximative contracted quasipotential $\phi_{\tilde{m}}(m)$ with
its two minima. Part b) contains two histograms for
$-\frac{1}{N}\log\hat{w}_{\tilde{m}}(m)$, the lower one with $N=10^{5}$ and
the upper one with $N=10^{6}$. Part c) and d) show enlargements of these
histograms in the neighbourhood of the left and right minimum, respectively,
together with shifted portions of the contracted quasipotential (as solid
lines) in order to demonstrate that again the approximate contracted
quasipotentials correctly predict the shape of the histograms within the
numerical accuracy.

As a final example we look at the system with $a=0.5$, $\kappa=0.1$ and 
$\alpha=7.92$. In this case the mean field map is chaotic on an interval.
We use this case as an example of how to apply time delay contracted
quasipotentials (\ref{eq:3-220}). Fig.~5 a) shows the result of a
simulation of this system with $N=3000$ in the form of a time delay plot. The
time delay contracted quasipotential in Fig.~5 b), which again has
been obtained with the approximation (\ref{eq:4-245}), allows predictions
about the blurring effect of finite $N$ in numerical simulations of such
systems. 

\section{Conclusion and outlook}
\label{se:5}

The aim of this article was to develop a method for describing large
deviations from the thermodynamic limit of systems of globally coupled noisy 
maps after the model of the quasipotential method for stochastically perturbed
dynamical systems. 

On an abstract level this has been achieved by interpreting the
time evolution in the thermodynamic limit as a consequence of a deterministic
dynamical system on a space of probability measures (Theorem \ref{th:1}) and
by recognising that the deviations from this thermodynamic limit behaviour in
one time step
follow a statistical distribution which allows exponential asymptotic
estimates for large deviations (Theorem \ref{th:2}). These are the two
ingredients necessary for constructing abstract quasipotentials on the space
of probability measures (Theorem \ref{th:3}).

These abstract quasipotentials contain the full information about large
deviations from the thermodynamic limit in the long term. However, being
defined on an infinite dimensional space of probability measures, they are 
rather inaccessible through concrete calculations and ill-suited for 
directly interpreting the observed long-term distribution of
macroscopic quantities in large finite systems.

Therefore, the concept of contracted quasipotentials has been introduced
(Theorem \ref{th:4} and equation (\ref{eq:3-220})) in order to extract from
the abstract quasipotential information about concrete and observable
characteristics of the temporal evolution of macroscopic quantities.

The Hamiltonian approach discussed in Section \ref{se:4} makes it clear how
difficult it is to solve the variational problem behind the
quasipotentials . On the other hand it opens the way for
approximations which turn the problem into a manageable one. In particular, the
approximation (\ref{eq:4-290}) is appropriate in the vicinity of attractors of
the macroscopic dynamics. 

The simple example with linearly and nonlinearly
coupled linear maps shows that this approximation gives a correct description
of the probability of rare deviations from the thermodynamic limit close to
an attractor. An investigation of the much more interesting situation of
coupled nonlinear maps like the tent map or the logistic map, which requires
the implementation of a new numerical scheme, is in progress and is expected
to lead to a correct description of the blurring effect of finite system size
on the attractors of the macroscopic dynamics in such systems.
In addition, this approach will offer a new answer to the question under which
circumstances the invariant density, which is a fixed point of the nonlinear
Frobenius-Perron equation, can loose stability. It will be interesting to
compare this new criterion to the various statements obtained by linear
stability analysis of the Frobenius Perron dynamics, which depend on
which norm is used for measuring the smallness of perturbations
\cite{Jus1,Jus2,Jus3,ErP1,ErP2,Mor,CMor,NKpp}. 

However, as the approximation (\ref{eq:4-290}) is limited to the vicinity of
attractors, it is desirable to go beyond the approximation (\ref{eq:4-245}) of
equation (\ref{eq:4-130}) so that it is possible to calculate global features
of quasipotentials like quasipotential barriers between different attractors.
This is probably not possible for the general case, but for special choices of
$\mathcal{F}$ there is some hope of being able to simplify equation
(\ref{eq:4-130}) --- for instance if the maps $F_{\kappa}(x,\mu)$ are
piecewise linear in $x$ and the noise influence in $\mathcal{F}$ is weak
(i. e., if $Q_{x}$ is close to $\delta_{x}$).

In view of this problematic situation with the evaluation of quasipotentials 
it is consoling, that there is a class of systems for which the computation of
quasipotentials for finite-$N$ effects in globally coupled systems is not more
difficult than the computation of quasipotentials for weak noise effects in
low-dimensional dynamical systems. This class of systems consists of coupled
subsystems which are defined on a finite state space $M$. More concretely, if
$M$ consists of $m$ elements, empirical measures can be written as
$(m-1)$-dimensional vectors, so that $\mathcal{F}$ defines a
$(m-1)$-dimensional dynamical system, and the quasipotential $\Phi$ is defined
on a $(m-1)$-dimensional space. While such systems with a finite discrete
subsystem space are not directly motivated by realistic physical systems,
they can be studied as conceptually simplest prototypes for globally coupled
noisy systems that show non-trivial effects such as noise induced phase
transitions when $N$ is finite, which can be analysed using quasipotentials.
Results in this direction will be published elsewhere \cite{HmDcr}. 

\section*{Acknowledgement}
Support by the {\em Deutsche Forschungsgemeinschaft} through SFB~237 is
gratefully acknowledged.

\newpage
\begin{appendix}

\section{Appendix}
This Appendix gives details of a proof of Theorem \ref{th:2}. The reader who
is not familiar with the mathematics literature on {\em large deviations} will
find this theorem less intuitive than most other statements in this article,
and therefore --- instead of referring to related results in \cite{DawG} --- we
will try to use a fairly self-contained line of arguments here, following the
proof of Sanov's theorem in \cite{EllD}, but without making explicit use
of stochastic control theory.

We try to use a language which keeps close to the standard conventions of
probability theory \cite{Dud,Tay}, 
but at the same time makes the formulae readable even
without a full technical understanding of mathematical details. In particular,
all probabilities about which we talk are derived from a probability measure
$\Prob:\mathcal{A}\to\Rset_{0}^{+}$ on a measure space $(\Omega,\mathcal{A})$,
where $\mathcal{A}$ is a sigma-algebra of events, which are subsets of
$\Omega$. A random variable $R$ with state space $(M,\mathcal{B}_{M})$ (where
$\mathcal{B}_{M}$ is a sigma algebra of (Borel) subsets of $M$) is a
measurable function $R:\Omega\to M$, and $\Prob\{R\in A\}$ for $A\subset M$ is
a shorthand notation for $\Prob(\{\omega\in\Omega : R(\omega)\in A\})$. 
We write $\Prob\{R\in\cdot\}$ for the probability measure on $M$
induced by $R$, and integrals over functions $f:M\to\Rset$ with respect to
this measure as $\int_{M} f(m) \Prob\{R\in \d m\}$. If $R$ and $S$ are
two random variables on state spaces $M_{R}$ and $M_{S}$, then we write the
conditional probability that $R$ is in $A$ conditioned on $S=s$, $s\in M_{s}$,
as $\Prob\{R\in A|S=s \}$. This object can be interpreted as a
measurable function $s\mapsto \Prob\{R\in A|S=s \}$ with the property
\begin{equation}
\label{eq:A-10}
\int_{B}\Prob\{R\in A|S=s\} \Prob\{S\in \d s\} = \Prob\{R\in A \mbox{ and }
S\in B\}  
\end{equation}
(which is not unique since there are several functions with this property, but
they differ only on sets of $\Prob\{S\in\cdot\}$-measure zero). The
topological assumptions of this paper guarantee that for every fixed $s\in S$, 
$\Prob\{R\in\cdot|S=s\}$ is a probability measure.

Before starting with the proof it is useful to see an important property of
the relative entropy:

{\em
Let $g:M\to\Rset$ be a bounded measurable function and
$\theta\in\mathcal{P}(M)$. Then the following variational formula holds true:
\begin{equation}
\label{eq:A-20}
  -\log \int_{M} e^{-g(m)} \theta(\d m) =
  \inf_{\chi\in\mathcal{P}(M)} \left\{ H(\chi|\theta) +
                                  \int_{M} g(m) \chi(\d m) \right\} .
\end{equation}
}

In order to see that this is true, it is sufficient to check the right hand
side of (\ref{eq:A-20}) for measures $\chi$ which are absolutely continuous
with respect to $\theta$, since otherwise $H(\chi|\theta)=\infty$.

A density 
\begin{equation}
\label{eq:A-30}
  \frac{\d \chi_{0}}{\d \theta}(m) :=
  \frac{e^{-g(m)}}{\int_{M}e^{-g(y)}\theta(\d y)}
\end{equation}
defines a measure $\chi_{0}\in\mathcal{P}(M)$, and obviously a measure
$\chi$ which is absolutely continuous with respect to $\theta$ is
absolutely continuous with respect to $\chi_{0}$, too. Therefore we can
write (cf. eq. (\ref{eq:3-70})):
\[
  H(\chi|\theta) = \int_{M} \left( \log\frac{\d\chi}{\d\theta}
                                   \right) \d\chi    \\
                 = \int_{M} \left( \log\frac{\d\chi}{\d\chi_{0}}
                                   \right) \d\chi                
                 + \int_{M} \left( \log\frac{\d\chi_{0}}{\d\theta}
                                   \right) \d\chi ,               
\]
and this means because of \ref{eq:A-30}:
\begin{equation}
\label{eq:A-40}
  H(\chi|\theta) + \int_{M} g(m) \chi(\d m)
  =
  H(\chi|\chi_{0}) - \log \int_{M} e^{-g(m)}\theta(\d m) .  
\end{equation}
Since $H(\chi|\chi_{0})\geq 0$, with equality exactly for the case
$\chi=\chi_{0}$, the variational formula (\ref{eq:A-20}) is true.

Now we turn to the proof of Theorem \ref{th:2}. We introduce the following
abbreviation:
\begin{equation}
\label{eq:A-50}
  V_{N} := -\frac{1}{N} \log \int_{\mathcal{P}(M)}
           e^{-Nh(\varphi)} \Prob\{ 
                      \mathcal{E}^{[N]}_{\mathbf{X}^{[N]}_{t+1}}\in\d\varphi |
                      \mathcal{E}^{[N]}_{\mathbf{X}^{[N]}_{t}}=\nu^{[N]}\} . 
\end{equation}
With this notation, Theorem \ref{th:2} states (see eq.~(\ref{eq:3-40})):
\begin{equation}
\label{eq:A-60}
  \lim_{N\to\infty} V_{N} = \inf_{\varphi\in\mathcal{P}(M)}
  \left[ H(\varphi|\mathcal{F}(\nu)) + h(\varphi) \right] .
\end{equation}

In order to show this statement, we will first derive a different
representation of $V_{N}$.
The construction of the new representation starts with building up
the empirical measure $\mathcal{E}_{\mathbf{x}^{[N]}}^{[N]}$ step by step:
If we define
\begin{equation}
\label{eq:A-80}
  \mathcal{E}_{\mathbf{x}^{[N]}}^{[0]} := 0
\end{equation}
then the following iteration:
\begin{equation}
\label{eq:A-90}
  \mathcal{E}_{\mathbf{x}^{[N]}}^{[j+1]} :=
  \mathcal{E}_{\mathbf{x}^{[N]}}^{[j]} + \frac{1}{N}\delta_{x^{(j+1)}}  
\end{equation}
for $j=0,\dots,N-1$
leads to $\mathcal{E}_{\mathbf{x}^{[N]}}^{[N]}$.
Note that $\mathcal{E}_{\mathbf{x}^{[N]}}^{[j]}$ is an element of 
the space of measures $\psi$ on $M$ which have
the property $\psi(M)=\frac{j}{N}$, and this space we denote by
$\mathcal{P}_{\frac{j}{N}}(M)$.

Now we want to define an iterative scheme that leads to $V_{N}$. The
intermediate steps involve the following quantities (with $j=0,\dots,N$ and
$\psi\in\mathcal{P}_{\frac{j}{N}}(M)$): 
\begin{eqnarray}
  & & \hspace{-5mm} V_{N}^{[j|\psi]} := \nonumber\\
\label{eq:A-100}
  & & \hspace{-3mm} -\frac{1}{N} \log \int_{\mathcal{P}(M)}
                     e^{-Nh(\varphi)} \Prob\{ 
                      \mathcal{E}^{[N]}_{\mathbf{X}^{[N]}_{t+1}}\in\d\varphi |
                      \mathcal{E}^{[j]}_{\mathbf{X}^{[N]}_{t+1}}=\psi 
                      \mbox{ and }
                      \mathcal{E}^{[N]}_{\mathbf{X}^{[N]}_{t}}=\nu^{[N]}\} . 
\end{eqnarray} 
The two extreme cases of this definition are
\begin{equation}
\label{eq:A-110}
  V_{N}^{[0|0]} = V_{N}
\end{equation}
and
\begin{equation}
\label{eq:A-120}
  V_{N}^{[N|\varphi]} = h(\varphi)
\end{equation}
for $\varphi\in\mathcal{P}(M)$.

In search for an iterative connection between these two cases we write down
the following sequence of equations:
\begin{eqnarray*}
  e^{-NV_{N}^{[j|\psi]}} & & = 
      \int_{\mathcal{P}(M)} e^{-Nh(\varphi)} \Prob\{
                      \mathcal{E}^{[N]}_{\mathbf{X}^{[N]}_{t+1}}\in\d\varphi |
                      \mathcal{E}^{[j]}_{\mathbf{X}^{[N]}_{t+1}}=\psi 
                      \mbox{ and }
                      \mathcal{E}^{[N]}_{\mathbf{X}^{[N]}_{t}}=\nu^{[N]}\}\\
  & & = \int_{\mathcal{P}_{\frac{j+1}{N}}(M)}
        \left(
      \int_{\mathcal{P}(M)} e^{-Nh(\varphi)} \Prob\{
                 \mathcal{E}^{[N]}_{\mathbf{X}^{[N]}_{t+1}}\in\d\varphi |
                 \mathcal{E}^{[j+1]}_{\mathbf{X}^{[N]}_{t+1}}=\tilde{\varphi} 
                 \mbox{ and }
                 \mathcal{E}^{[N]}_{\mathbf{X}^{[N]}_{t}}=\nu^{[N]}\}
        \right)  \\
  & &            \hspace{30mm}               \Prob\{
            \mathcal{E}^{[j+1]}_{\mathbf{X}^{[N]}_{t+1}}\in\d\tilde{\varphi} |
            \mathcal{E}^{[j]}_{\mathbf{X}^{[N]}_{t+1}}=\psi 
            \mbox{ and }
            \mathcal{E}^{[N]}_{\mathbf{X}^{[N]}_{t}}=\nu^{[N]}\} \\
  & & = \int_{\mathcal{P}_{\frac{j+1}{N}}(M)}
            e^{-NV_{N}^{[j+1|\tilde{\varphi}]}}
            \Prob\{
            \mathcal{E}^{[j+1]}_{\mathbf{X}^{[N]}_{t+1}}\in\d\tilde{\varphi} |
            \mathcal{E}^{[j]}_{\mathbf{X}^{[N]}_{t+1}}=\psi 
            \mbox{ and }
            \mathcal{E}^{[N]}_{\mathbf{X}^{[N]}_{t}}=\nu^{[N]}\} \\
  & & = \int_{M} e^{-NV_{N}^{[j+1|\psi+\frac{1}{N}\delta_{y}]}}
            \Prob\{ X_{t+1}^{(j+1)}\in\d y | 
                 \mathcal{E}^{[N]}_{\mathbf{X}^{[N]}_{t}}=\nu^{[N]}\} \\
  & & = \int_{M} e^{-NV_{N}^{[j+1|\psi+\frac{1}{N}\delta_{y}]}}
            (\mathcal{F}(\nu^{[N]}))(\d y)
\end{eqnarray*}
where in the last step we have used eq. (\ref{eq:3-35}). If we take the
logarithm of this equation and apply the variational formula (\ref{eq:A-20})
with the settings 
$\theta=\mathcal{F}(\nu^{[N]})$ and
$g(m)=NV_{N}^{[j+1|\psi+\frac{1}{N}\delta_{m}]}$ 
then we obtain
\begin{equation}
\label{eq:A-130}
  V_{N}^{[j|\psi]} =
   \inf_{\varphi\in\mathcal{P}(M)} \left[
       \frac{1}{N} H(\varphi|\mathcal{F}(\nu^{[N]}))
              + \int_{M}V_{N}^{[j+1|\psi+\frac{1}{N}\delta_{y}]} \varphi(\d y)
       \right] .
\end{equation}
In the light of \cite{HmSem} this can be called a stochastic possibilistic
dynamics. 
If we start from $V_{N}^{[N|\psi]}=h(\psi)$ and iterate eq.~(\ref{eq:A-130})
$N$ times we arrive at the following equation:
\begin{eqnarray}
  & & \hspace{-5mm}V_{N} = \nonumber\\
\label{eq:A-140}
  & & \hspace{-5mm} 
       \inf_{\varphi_{1},\dots,\varphi_{N}\in\mathcal{P}(M)}
          \left\{
           \frac{1}{N} \sum_{i=1}^{N} H(\varphi_{i}|\mathcal{F}(\nu^{[N]}))
     + \int_{M}\dots\int_{M}h\left(\mathcal{E}_{\mathbf{y}^{[N]}}^{[N]}\right)
            \prod_{j=1}^{N} \varphi_{j}(\d y_{j})
          \right\} . 
\end{eqnarray}
This representation of $V_{N}$ is very convenient for showing half of the
statement (\ref{eq:A-60}) of Theorem \ref{th:2}, namely
\begin{equation}
\label{eq:A-150}
  \lim\sup_{N\to\infty}V_{N} \leq
  \inf_{\varphi\in\mathcal{P}(M)}
  [H(\varphi|\mathcal{F}(\nu)) + h(\varphi)] .
\end{equation} 
This inequality follows from eq. (\ref{eq:A-140}) in the following way:
It is a consequence of (\ref{eq:A-140}) that
\begin{equation}
\label{eq:A-160}
  V_{N} \leq \inf_{\varphi\in\mathcal{P}(M)}
             \left\{ H(\varphi|\mathcal{F}(\nu^{[N]}))
     + \int_{M}\dots\int_{M}h\left(\mathcal{E}_{\mathbf{y}^{[N]}}^{[N]}\right)
            \prod_{j=1}^{N} \varphi(\d y_{j})
          \right\} . 
\end{equation}
Using the facts that $\nu^{[N]}$ converges weakly to $\nu$ and
$\mathcal{E}_{\mathbf{Y}^{[N]}}^{[N]}$ converges weakly to $\varphi$ with
probability 1, this inequality leads directly to (\ref{eq:A-150}).

Now the only missing part in the proof of Theorem \ref{th:2} is the second
half of (\ref{eq:A-60}), namely
\begin{equation}
\label{eq:A-170}
  \lim\inf_{N\to\infty}V_{N} \geq
  \inf_{\varphi\in\mathcal{P}(M)}
  [H(\varphi|\mathcal{F}(\nu)) + h(\varphi)] .
\end{equation} 
Unfortunately this requires more work than (\ref{eq:A-150}), and we have to
use some non-trivial but standard results from probability theory which can
be found in introductory texts like \cite{Dud,Tay}.

First of all we observe that the convexity of $H$ allows us to use Jensen's
inequality in (\ref{eq:A-140}), so that
\begin{eqnarray}
  & & \hspace{-6mm}V_{N} \geq \nonumber\\
\label{eq:A-180}
  & & \hspace{-6mm} 
       \inf_{\varphi_{1},\dots,\varphi_{N}\in\mathcal{P}(M)}
         \left\{ H \left(
          \frac{1}{N} \sum_{i=1}^{N} \varphi_{i}|\mathcal{F}(\nu^{[N]})\right)
     + \int_{M}\dots\int_{M}h\left(\mathcal{E}_{\mathbf{y}^{[N]}}^{[N]}\right)
            \prod_{j=1}^{N} \varphi_{j}(\d y_{j})
          \right\} . 
\end{eqnarray}
Introducing the abbreviation
\begin{equation}
\label{eq:A-190}
  \alpha_{N} := \frac{1}{N} \sum_{i=1}^{N} \varphi_{i} 
\end{equation}
we can rewrite the content of (\ref{eq:A-180}):

For every $\varepsilon>0$ and every $N\in\Nset$ there is a sequence
$(\varphi_{1},\dots,\varphi_{N})$ of elements in $\mathcal{P}(M)$ such that
\begin{equation}
\label{eq:A-200}
  V_{N}+\varepsilon \geq
     \int_{M}\dots\int_{M}\left[H(\alpha_{N}|\mathcal{F}(\nu^{[N]})) +
       h\left(\mathcal{E}_{\mathbf{y}^{[N]}}^{[N]}\right) 
            \right] \prod_{j=1}^{N} \varphi_{j}(\d y_{j}) .
\end{equation}

We are going to establish a connection between the measures $\alpha_{N}$ in
the first term and the empirical measures in the second term of the above
sum:

Let $Y^{(i)}$ be a random variable with values in $M$ and
$\Prob(Y^{(i)}\in A)=\varphi_{i}(A)$. If $g:M\to\Rset$ is bounded and
continuous, then we find that the sequence of differences
\begin{equation}
\label{eq:A-210}
  \int_{M}g(x)\mathcal{E}_{\mathbf{Y}^{[N]}}^{[N]}(\d x) -
  \int_{M}g(x)\alpha_{N}(\d x) =
  \frac{1}{N}\sum_{j=1}^{N} 
    \left[ g(Y^{(j)}) - \int_{M}g(x)\varphi_{j}(\d x) \right]  
\end{equation}
converges to $0$ with probability $1$ as $N\to\infty$, due to the strong law of
large numbers for non-identically distributed orthogonal random variables
\cite{Tay}. 

Now consider an arbitrary infinite subsequence $(\alpha_{N_{i}})$ of the
sequence $(\alpha_{N})$. Since $\mathcal{P}(M)$ is compact, the subsequence
has a subsubsequence $(\alpha_{N_{i_{k}}})$ that converges to some
$\alpha_{*}\in\mathcal{P}(M)$. For notational simplicity we write
$\alpha_{N_{i_{k}}}=:\alpha_{(k)}$. The convergence of the difference
(\ref{eq:A-210}) tells us then, that with probability $1$
\begin{equation}
\label{eq:A-220}
  \int_{M}g(x)\mathcal{E}_{\mathbf{Y}^{[(k)]}}^{[(k)]}(\d x)
  \to \int_{M}g(x)\alpha_{*}(\d x) .  
\end{equation}
So far the almost sure convergence in (\ref{eq:A-220}) holds true only after
fixing $g$. In order to conclude from (\ref{eq:A-220}) that 
$\mathcal{E}_{\mathbf{Y}^{[(k)]}}^{[(k)]}$ converges weakly to $\alpha_{*}$
with probability 1, it is sufficient to know that there is a countable dense
subset of the space of bounded, uniformly continuous functions (with regard to
an appropriate metric), but this follows from a standard separability result,
see for instance Corollary 11.2.5 in \cite{Dud}.

Combining these convergence results with the continuity of $h$ and the lower
semicontinuity of $H$ we know now that almost surely
\begin{equation}
\label{eq:A-230}
  \lim\inf_{k\to\infty}
  \left[H(\alpha_{(k)}|\mathcal{F}(\nu^{[(k)]})) +
       h\left(\mathcal{E}_{\mathbf{Y}^{[(k)]}}^{[(k)]}\right) 
            \right]  
  \geq
  \left[H(\alpha_{*}|\mathcal{F}(\nu)) +
       h\left(\alpha_{*}\right) 
            \right] .
\end{equation}

Making use of Fatou's lemma this means together with (\ref{eq:A-200}) that for
all $\varepsilon>0$:
\begin{equation}
\label{eq:A-240}
  \lim\inf_{k\to\infty} V_{(k)} + \varepsilon \geq
  \left[H(\alpha_{*}|\mathcal{F}(\nu)) +
       h\left(\alpha_{*}\right) 
            \right] \geq
  \inf_{\varphi\in\mathcal{P}(M)} \left[ H(\varphi|\mathcal{F}(\nu))
       + h(\varphi) \right] .        
\end{equation}

This is nearly the statement (\ref{eq:A-170}) except that it refers not to the
whole sequence $V_{N}$ but to a subsubsequence $V_{(k)}$. However, it has been
shown that every subsequence of $V_{N}$ has such a subsubsequence. Now assume
that eq.~(\ref{eq:A-170}) were not true. Then for sufficiently small
$\epsilon>0$ there would be an infinite sequence of $V_{N_{i}}$ smaller than
$ \inf_{\varphi\in\mathcal{P}(M)} 
\left[ H(\varphi|\mathcal{F}(\nu)) + h(\varphi) \right] - \epsilon$, and this
subsequence would not have a subsubsequence which fulfils
eq.~(\ref{eq:A-240}). This contradiction shows that (\ref{eq:A-170}) must be
true, and the proof of Theorem \ref{th:2} is complete.  

\end{appendix}

\newpage

\section*{Figure captions}
\begin{description}
\item[Fig. 1:] Time delay plot of the mean field $h_{t+1}=h[\rho_{t+1}]$ 
               versus $h_{t}=h[\rho_{t}]$
               as obtained by numerically solving the nonlinear
               Frobenius-Perron equation (\ref{eq:1-40}) for coupled tent
               maps (\ref{eq:1-20}) with $a=1.65$ and $\kappa=0.27$. 
\item[Fig. 2:] Time delay plot of the mean field $h_{t+1}$ 
               versus $h_{t}$ for the same parameters as Fig.~1, 
               but this time obtained by
               iterating equation (\ref{eq:1-10}) and using equation
               (\ref{eq:1-30}) for finite systems with decreasing size:
               {\bf a)} $N=10^{6}$, {\bf b)} $N=10^{5}$,  {\bf c)} $N=10^{4}$.
               In each case the plots show 2000 points after initial
               transients have died out.
\item[Fig. 3:] The contracted quasipotential $\phi_{\tilde{m}}(m)$ of the
               system (\ref{eq:4b-20}), (\ref{eq:4b-30})
               with linear coupling, compared to
               histogramms of the invariant density $\hat{w}_{\tilde{m}}(m)$
               obtained from numerical simulations: 
               {\bf a)} with parameters $\kappa=0$, $a=0$, $\sigma=1$,
               $N=10^{5}$ over $10000$ time steps,
               {\bf b)} with parameters $\kappa=0.4$, $a=0.5$, $\sigma=1$,
               and with $N=10^{5}$ (lower part) and
               $N=10^{6}$ over $10000$ time steps.
               In all cases the logarithmically plotted histogramms can be
               approximated by the contracted quasipotential (solid line)
               after a vertical shift by a constant offset.
\item[Fig. 4:] Comparison of invariant density and contracted quasipotential
               for the system (\ref{eq:4b-20}), (\ref{eq:4b-30}) 
               with nonlinear coupling 
               (\ref{eq:4b-110}). The parameters are $\kappa=0.1$, $a=0.5$, and
               $\alpha=3.16$; in this case, the nonlinear Frobenius-Perron
               dynamics predicts a period-2-attractor for the mean value.
               The various parts show {\bf a)} the contracted quasipotential
               $\phi_{\tilde{m}}(m)$, {\bf b)} a logarithmic plot of the
               histogramm for the invariant density $\hat{w}_{\tilde{m}}(m)$
               from a simulation of $10000$ time steps with $N=10^{5}$ 
               (lower histogramm) and
               $N=10^{6}$ (upper histogramm). Parts {\bf c)} and {\bf d)}
               enlarge the vicinity of the left periodic point and the right
               periodic point, respectively, to show the agreement between the
               histogramms and the shifted contracted quasipotials (solid
               lines). 
\item[Fig. 5:] {\bf a)} Time delay map, $m_{t+1}$ versus $m_{t}$ for the
               system (\ref{eq:4b-20}), (\ref{eq:4b-20}) 
               with nonlinear coupling 
               (\ref{eq:4b-110}) and parameters $\kappa=0.1$, $a=0.5$, and
               $\alpha=7.92$, obtained from a simulation with $N=3000$ over
               $10000$ time steps.
               {\bf b)} Time delay contracted quasipotential
               $\phi_{\tilde{m},2}(m,m')$. The contour lines for the levels of
               0.002, 0.004, and 0.006 in the quasipotential follow the shape
               of the blurred attractor in a). 
\end{description}

\newpage

\resizebox{14cm}{!}{\includegraphics{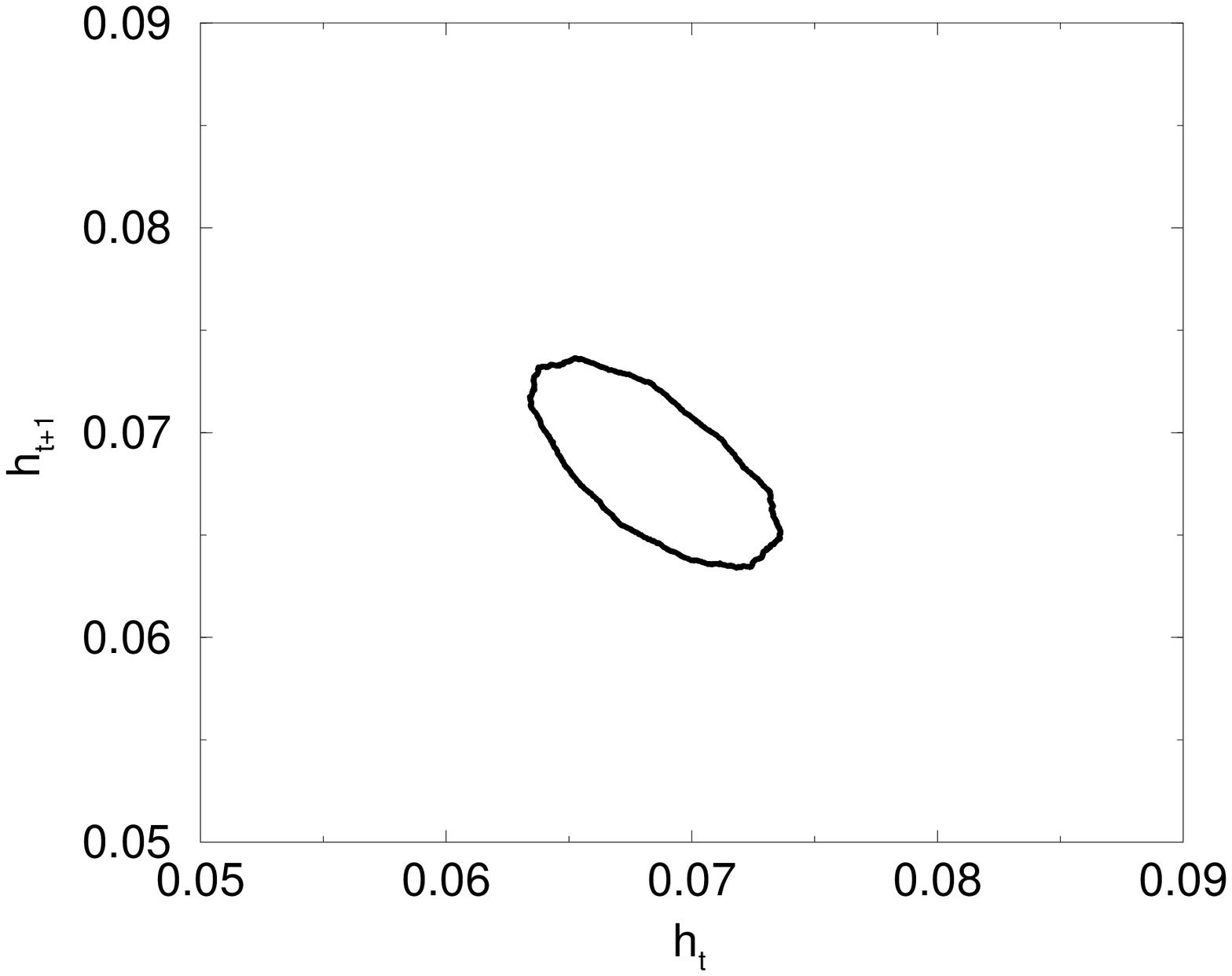}}
\vfill
\Large Fig. 1
\newpage
\resizebox{14cm}{!}{\includegraphics{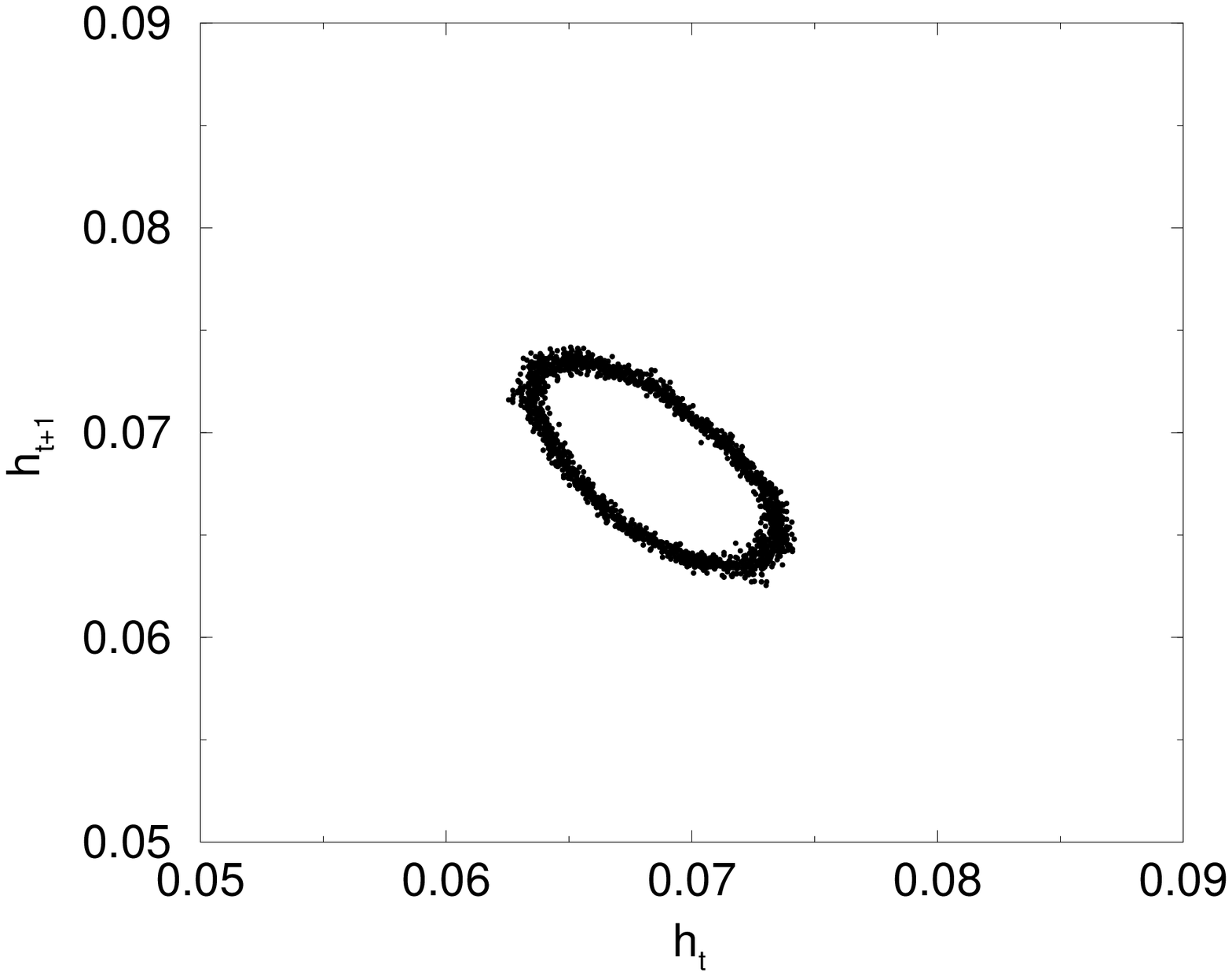}}
\vfill
\Large Fig. 2 a)
\newpage
\resizebox{14cm}{!}{\includegraphics{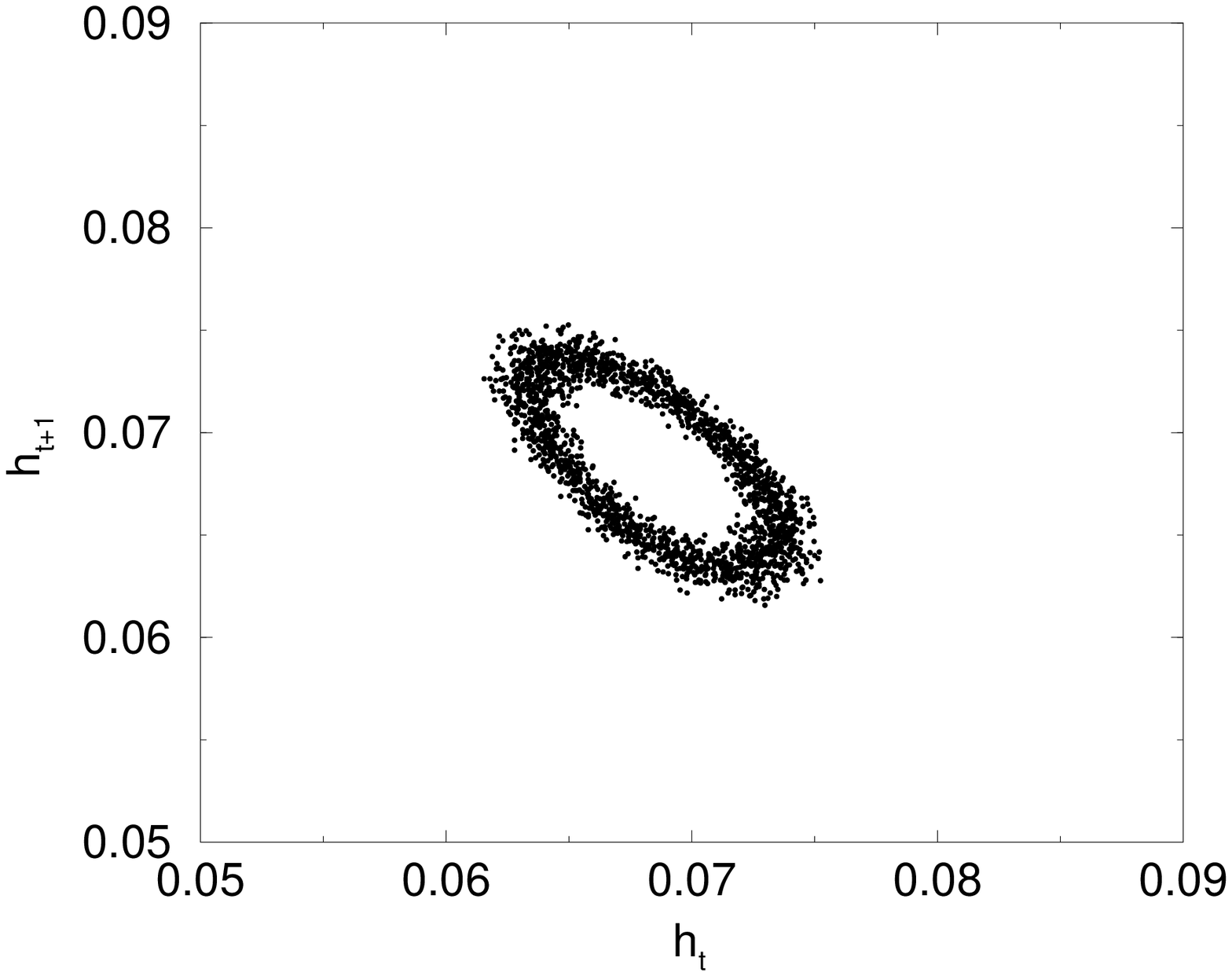}}
\vfill
\Large Fig. 2 b)
\newpage
\resizebox{14cm}{!}{\includegraphics{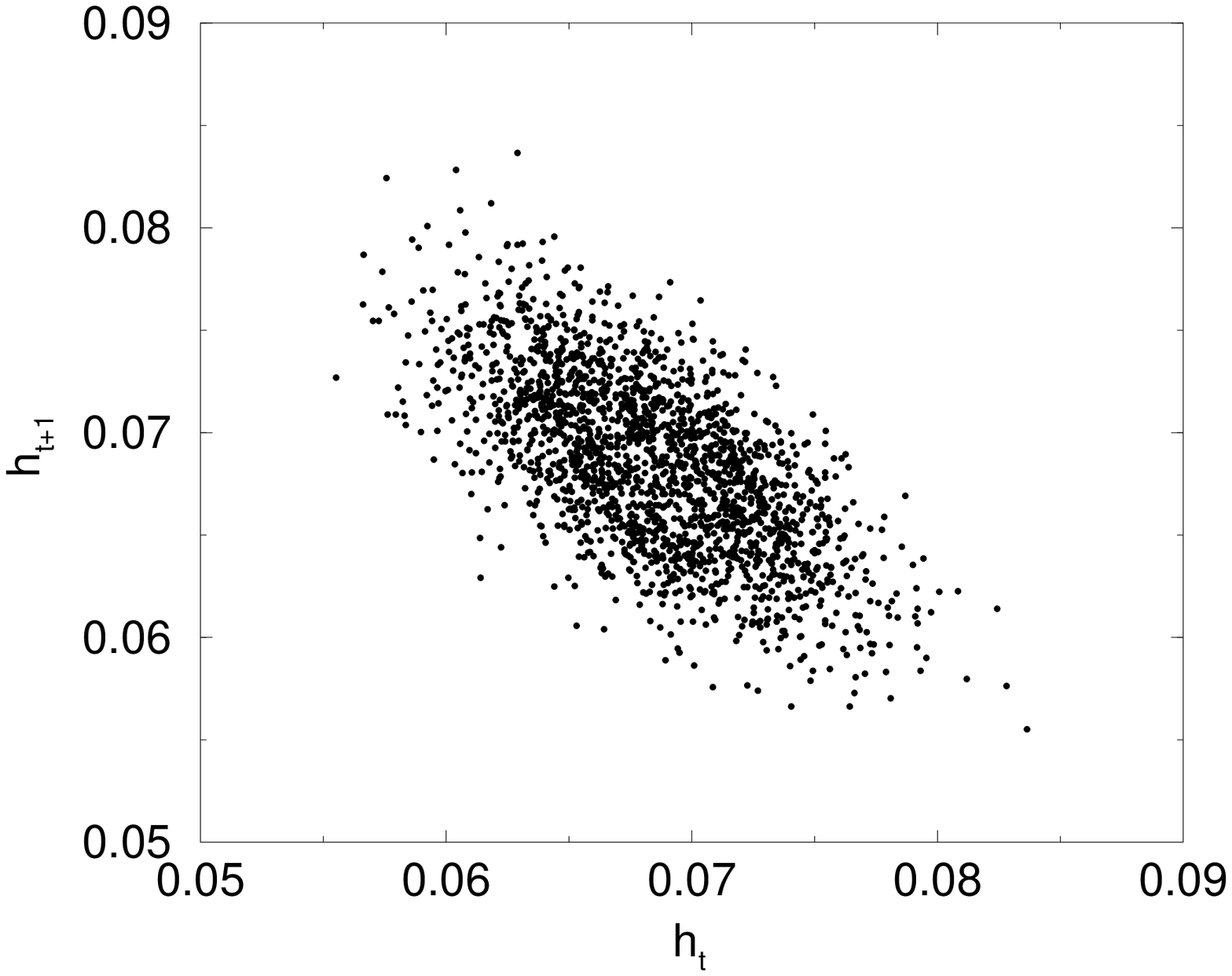}}
\vfill
\Large Fig. 2 c)
\newpage
\resizebox{14cm}{!}{\includegraphics{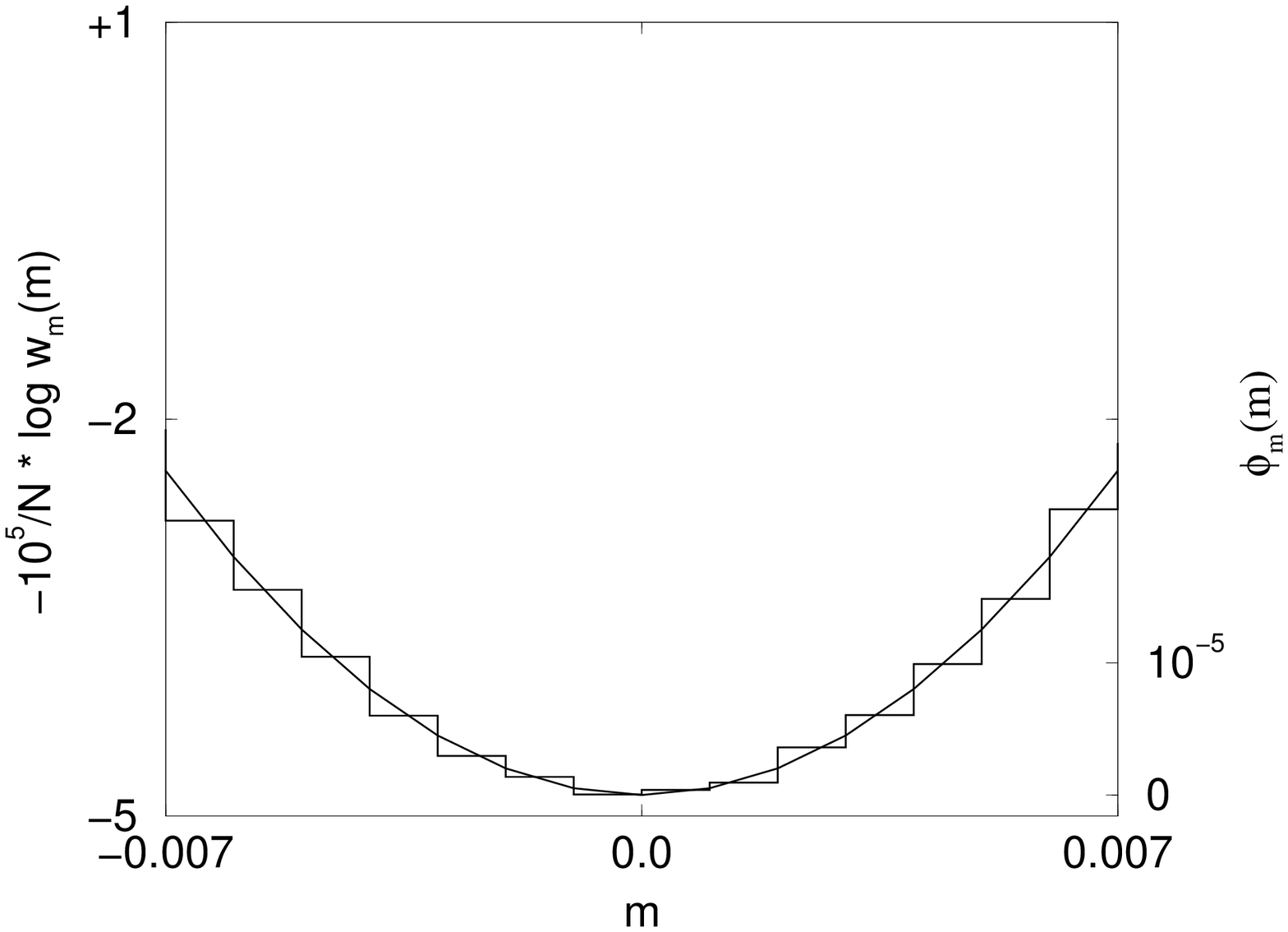}}
\vfill
\Large Fig. 3 a)
\newpage
\resizebox{14cm}{!}{\includegraphics{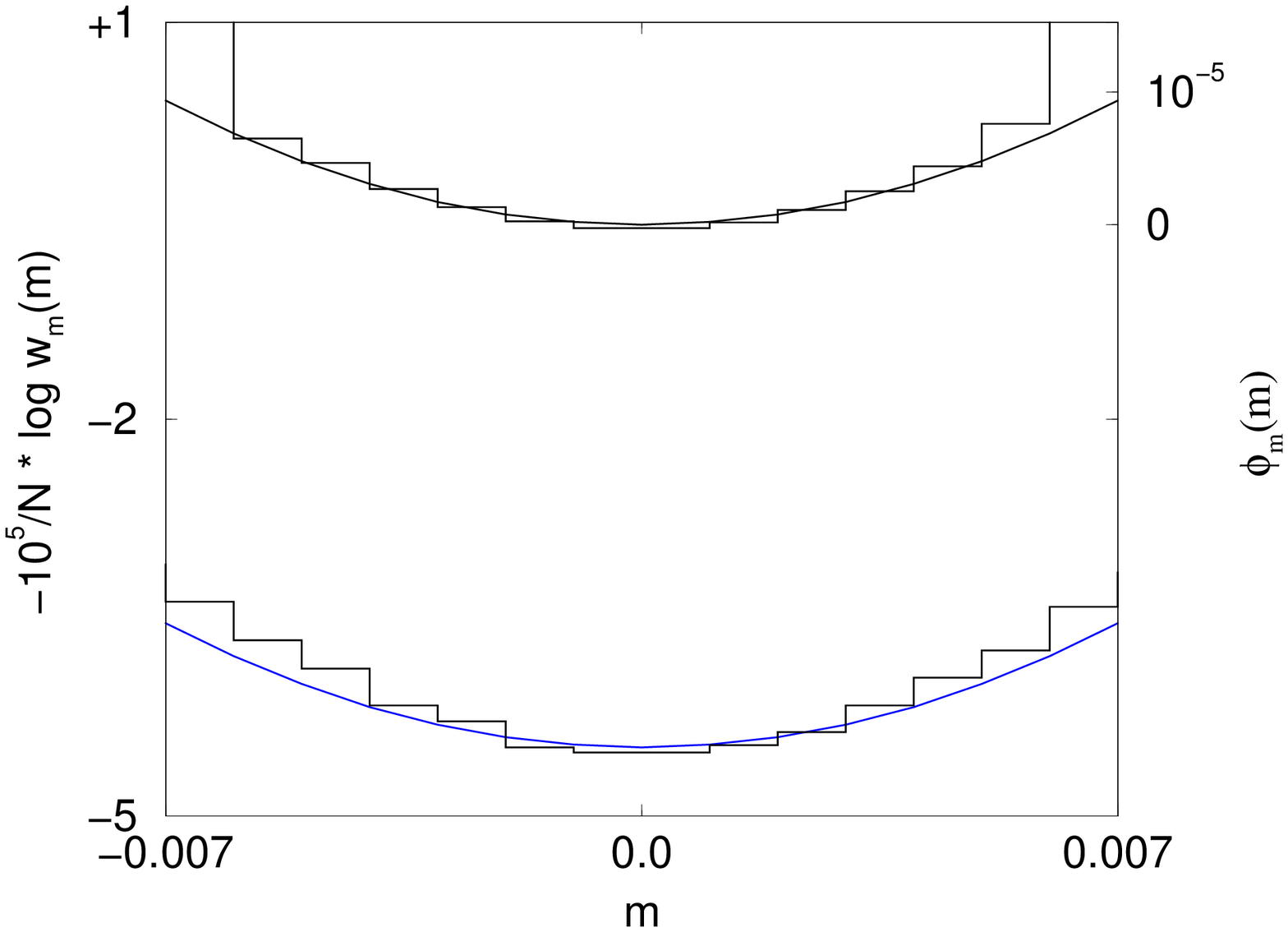}}
\vfill
\Large Fig. 3 b)
\newpage
\resizebox{14cm}{!}{\includegraphics{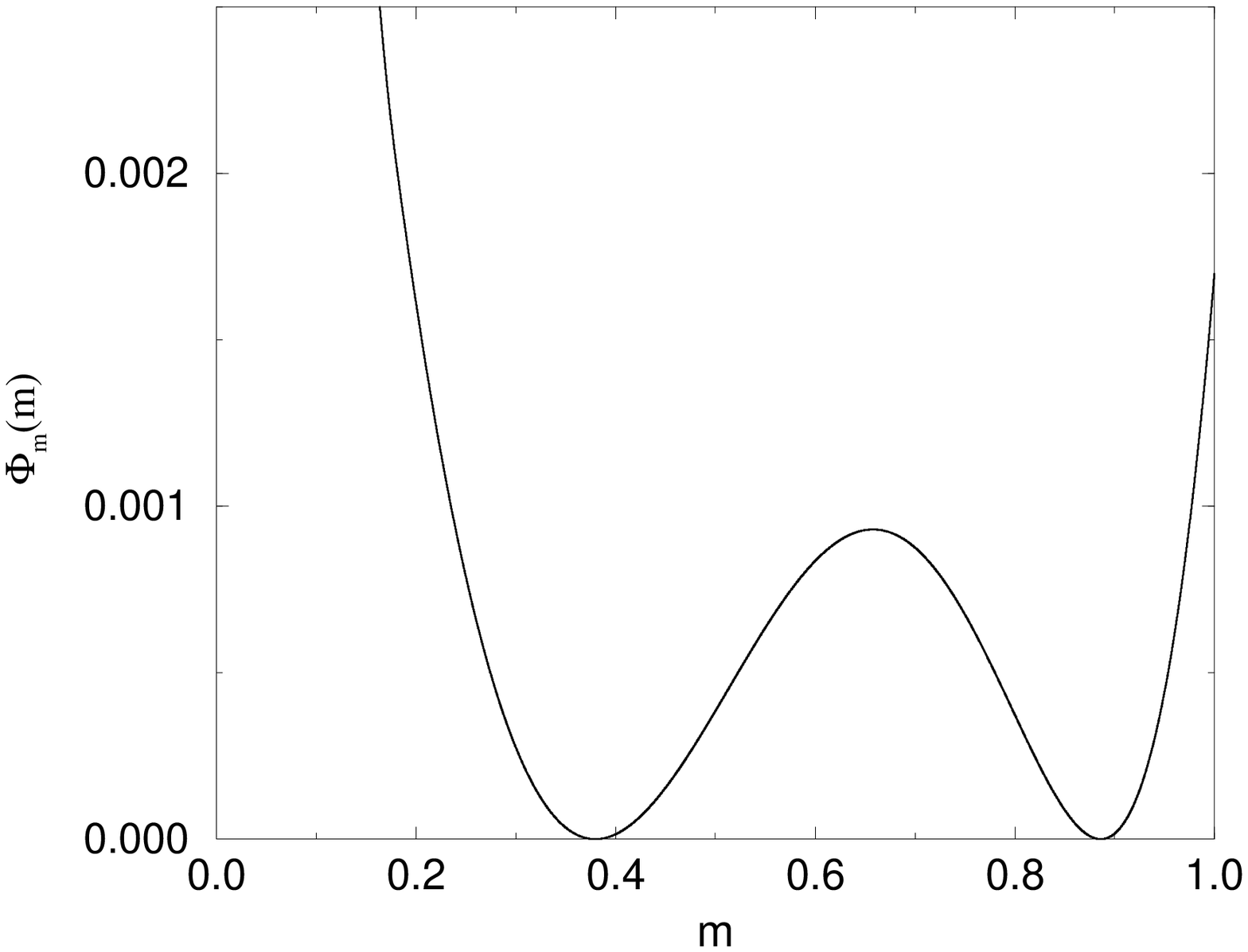}}
\vfill
\Large Fig. 4 a)
\newpage
\resizebox{14cm}{!}{\includegraphics{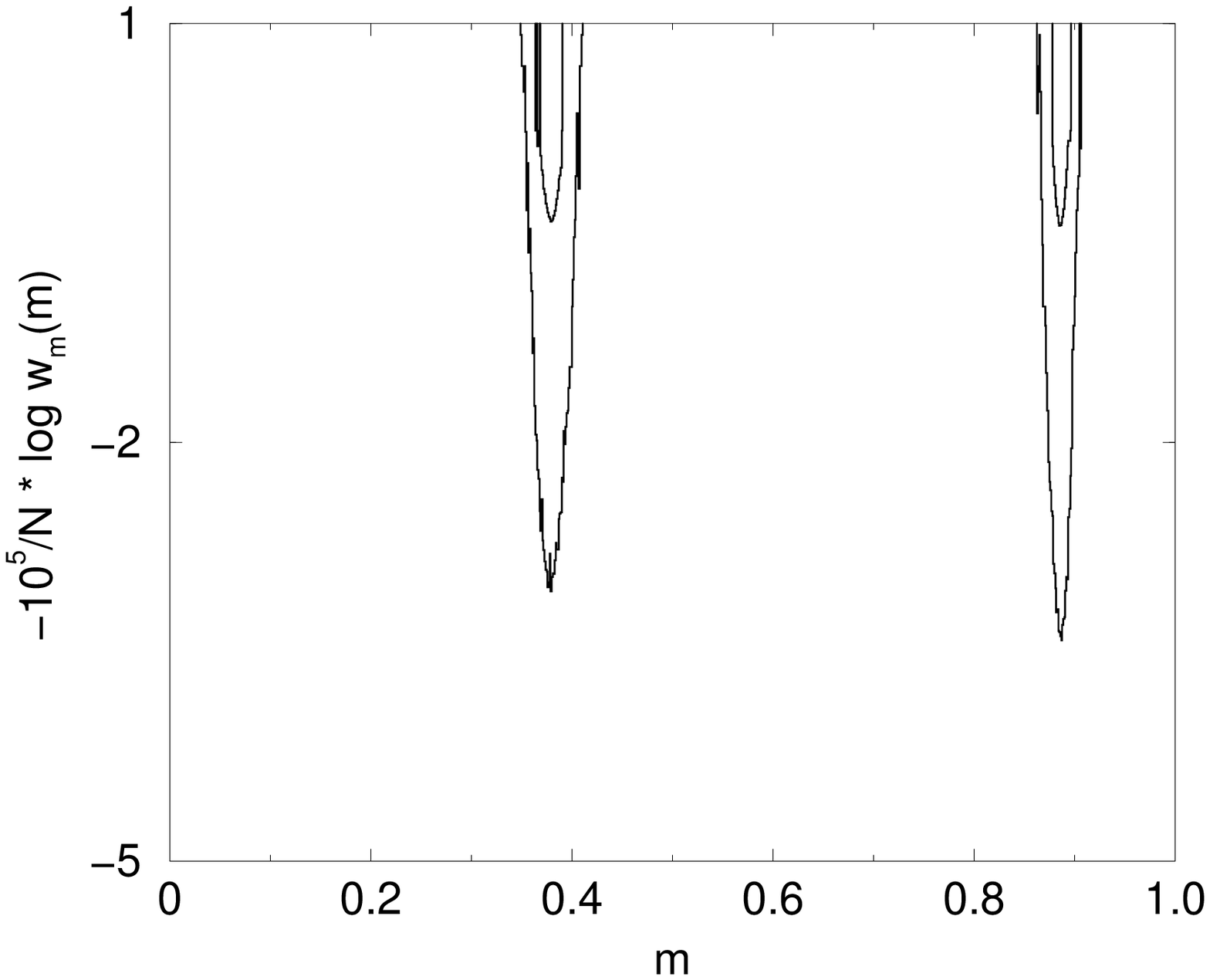}}
\vfill
\Large Fig. 4 b)
\newpage
\resizebox{14cm}{!}{\includegraphics{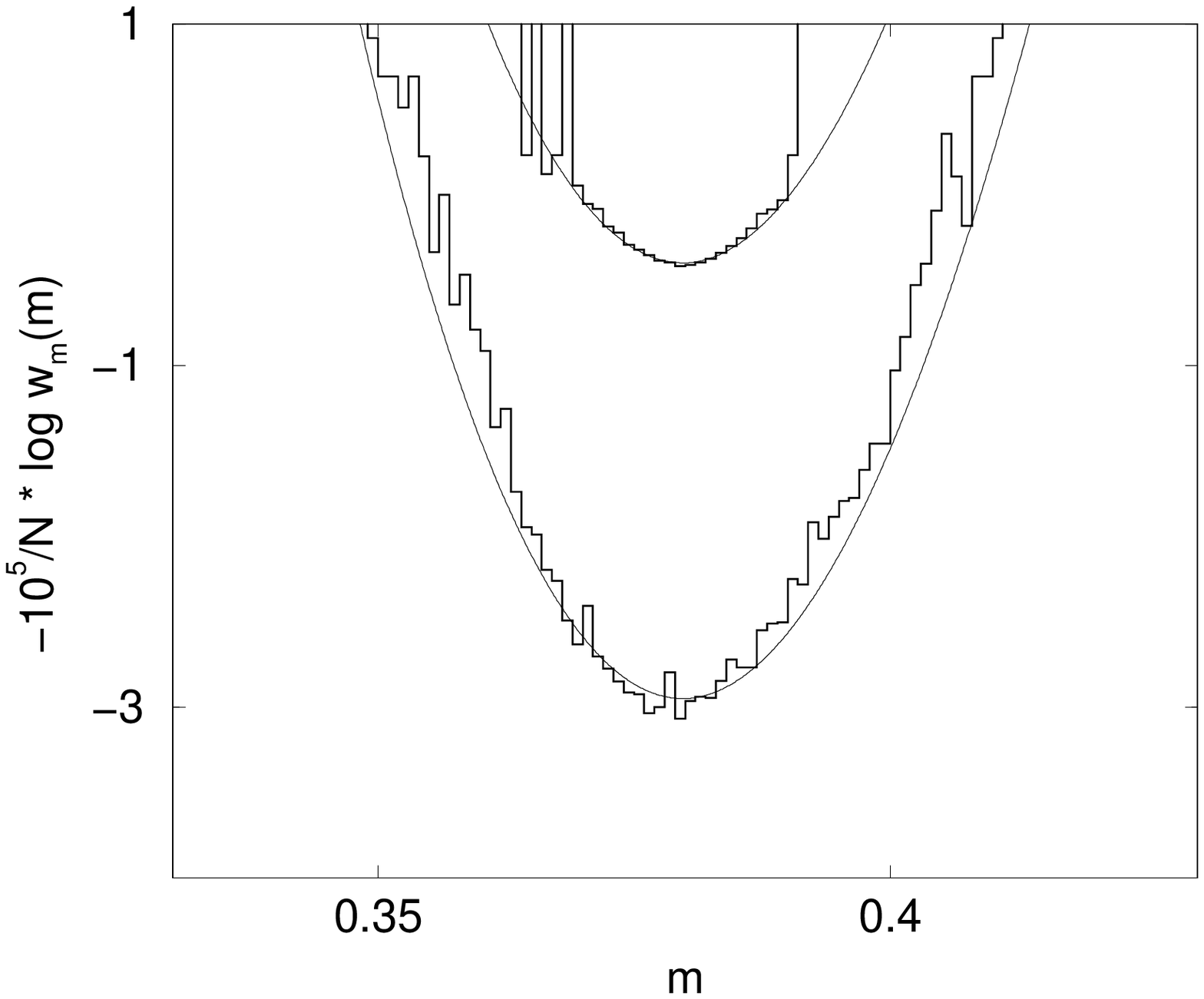}}
\vfill
\Large Fig. 4 c)
\newpage
\resizebox{14cm}{!}{\includegraphics{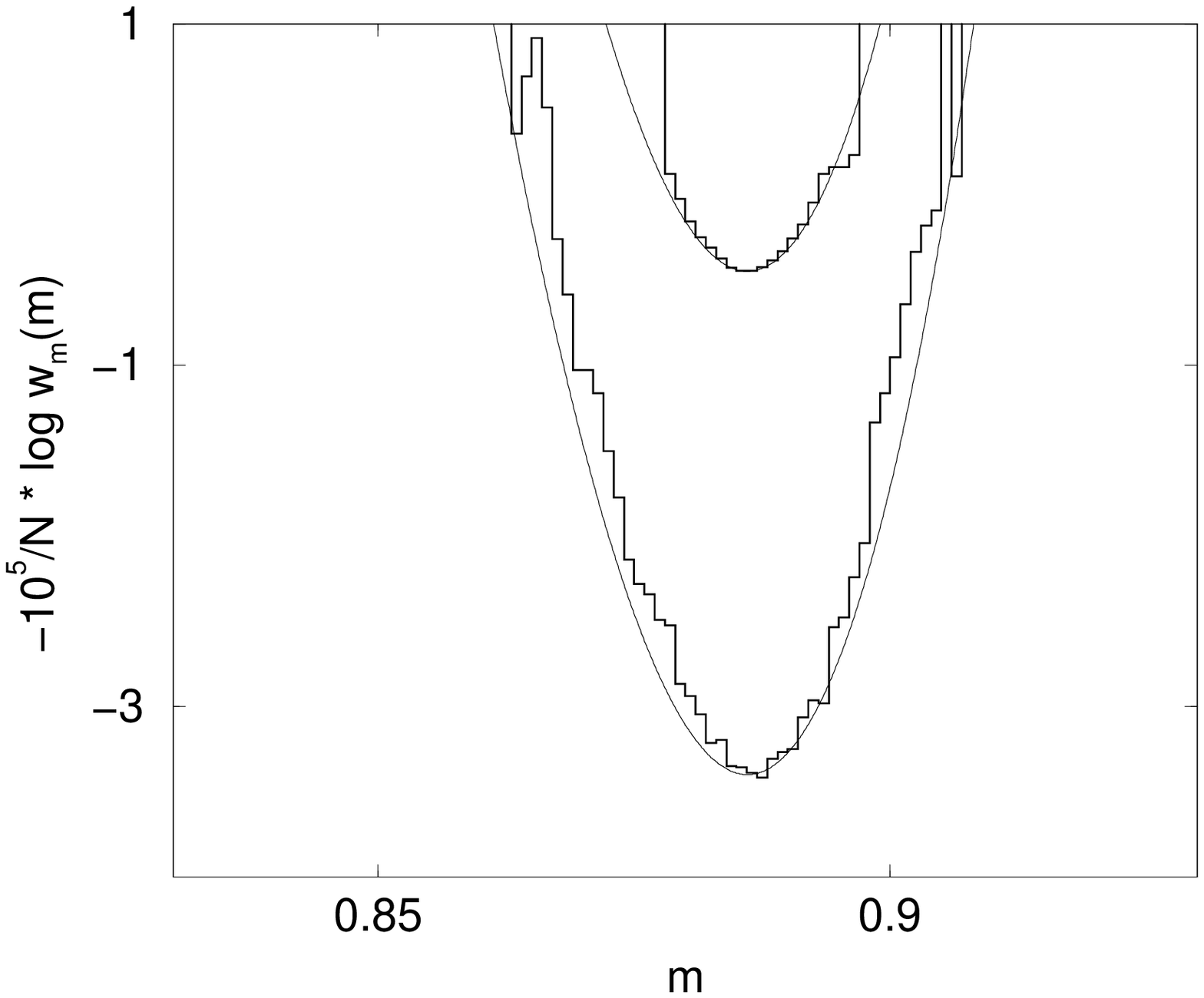}}
\vfill
\Large Fig. 4 d)
\newpage
\resizebox{14cm}{!}{\includegraphics{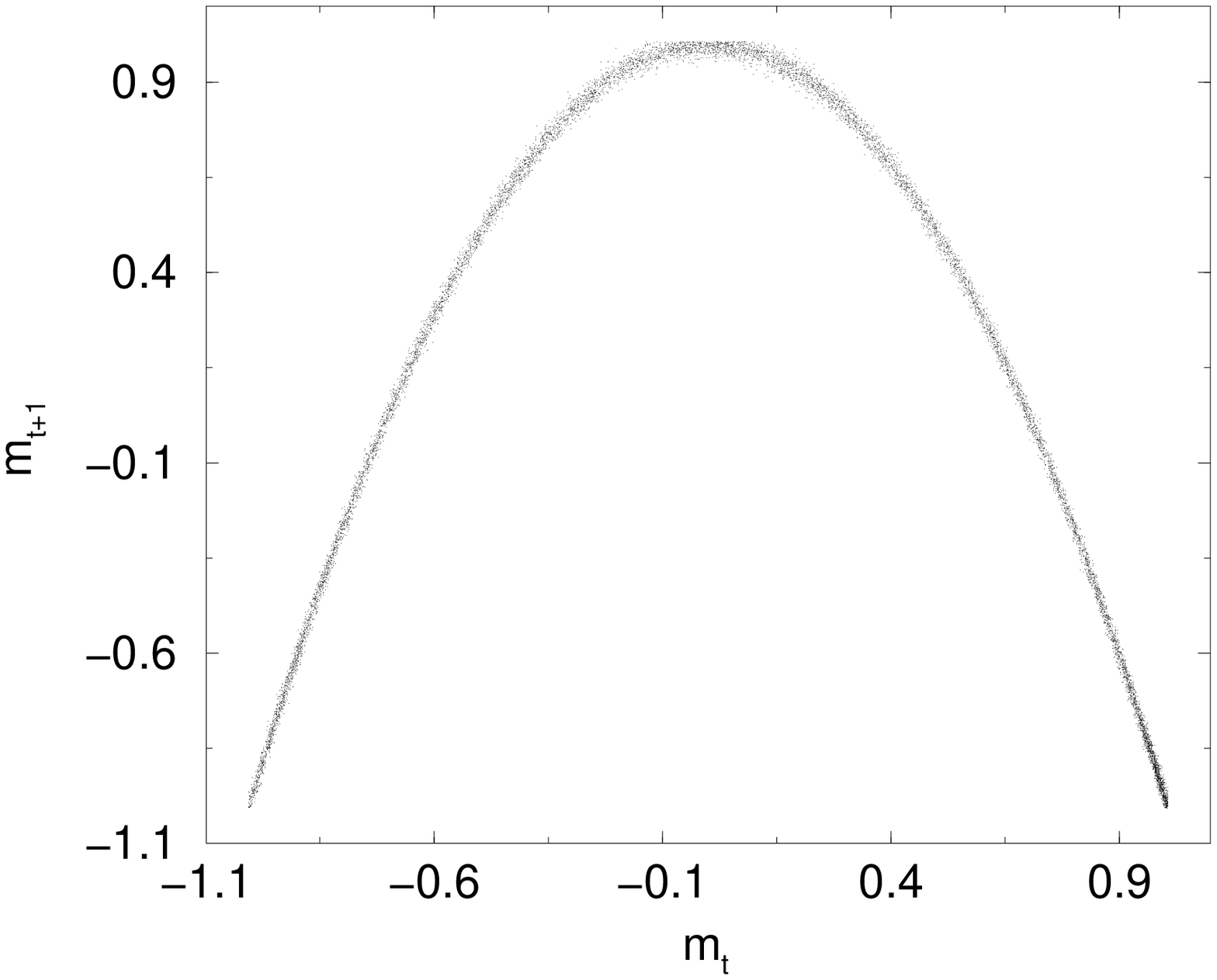}}
\vfill
\Large Fig. 5 a)
\newpage
\resizebox{14cm}{!}{\includegraphics{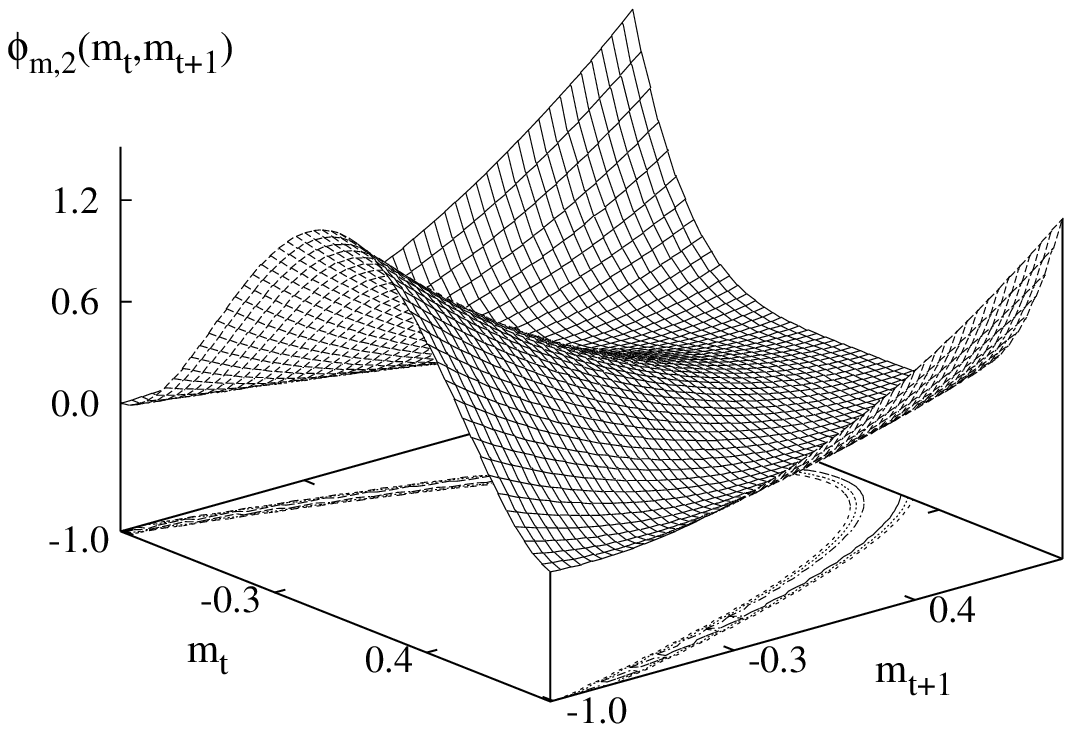}}
\vfill
\Large Fig. 5 b)
\end{document}